\documentclass[prd,twocolumn,floatfix,nopacs,preprintnumbers,nofootinbib]{revtex4}
\usepackage[utf8x]{inputenc}
\usepackage{mathrsfs}
\usepackage{amssymb}
\usepackage{amsmath}
\usepackage{mathtools}
\usepackage{slashed}
\usepackage{graphicx}
\usepackage{physics}

\usepackage[caption = false]{subfig}
\usepackage[normalem]{ulem}
\usepackage{xcolor}
\definecolor{lcolor}{rgb}{0.5,0,0}
\definecolor{citcolor}{rgb}{0,0.3,0.0}

\newlength{\mycol}
\usepackage[breaklinks,colorlinks,urlcolor=blue,citecolor=citcolor,linkcolor=lcolor]{hyperref}
\usepackage{mciteplus}
\newcommand{\acal}{{\mathcal{A}}}
\newcommand{\kcal}{{\mathcal{K}}}

\newcommand{\rt}{{\mathbf{r}}}
\newcommand{\xt}{{\mathbf{x}}}
\newcommand{\bt}{{\mathbf{b}}}

\newcommand{\nc}{N_\mathrm{c}}

\newcommand{\xpom}{{x_\mathbb{P}}}
\newcommand{\zmin}{z_{\text{min}}}
\newcommand{\Ydip}{Y_\text{dip}}
\newcommand{\Yqqg}{Y_{\text{q}\bar{\text{q}}\text{g}}}
\newcommand{\nlodip}{\mathrm{NLO}_\text{dip}}
\newcommand{\nloqqg}{\mathrm{NLO}_{\text{q}\bar{\text{q}}\text{g}}}
\newcommand{\lo}{\mathrm{LO}}
\newcommand{\nlo}{\mathrm{NLO}}
\newcommand{\li}{\mathrm{Li}_2}

\newcommand{\Ybk}{{Y_{0,\text{BK}}}}
\newcommand{\OLDME}{\langle \mathcal{O}_1\rangle_V}
\newcommand{\qLDME}{\langle \vec q^{\,2} \rangle_V}

\newcommand{\B}{/(\xt_{01}^2 e^{\gamma_E})}

\newcommand{\gev}{\ \textrm{GeV}}

\newcommand{\qs}{Q_\mathrm{s}}

\newcommand{\fig}{Fig.~}
\newcommand{\figs}{Figs.~}

\newcommand{\xbj}{{x_\text{bj}}}

\newcommand{\vcal}{\mathcal{V}}
\newcommand{\ical}{\mathcal{I}}

\newcommand{\as}{\alpha_\mathrm{s}}
\newcommand{\der}{\mathrm{d}}

\newcommand{\Deltat}{\mathbf{\Delta}}

\newcommand{\jpsi}{$\mathrm{J}/\psi$ }

\begin{document}
\author{Heikki Mäntysaari}
\email{heikki.mantysaari@jyu.fi}
\author{Jani Penttala}
\email{jani.j.penttala@student.jyu.fi}
\affiliation{
Department of Physics, University of Jyväskylä %
 P.O. Box 35, 40014 University of Jyväskylä, Finland
}
\affiliation{
Helsinki Institute of Physics, P.O. Box 64, 00014 University of Helsinki, Finland
}

\title{
Exclusive heavy vector meson production at next-to-leading order in the dipole picture
}

%\pacs{}

\preprint{}

\begin{abstract}
We calculate exclusive production of a longitudinally polarized heavy vector meson  at next-to-leading order in the dipole picture. The large quark mass allows us to separately include both the first QCD correction proportional to the coupling constant $\as$, and the first relativistic correction suppressed by the quark velocity $v^2$. Both of these corrections are found to be numerically important in \jpsi production. The results obtained are directly suitable for phenomenological calculations. We also demonstrate how vector meson production provides complementary information to structure function analyses when one extracts the initial condition for the energy evolution of the proton small-$x$ structure.

\end{abstract}

\maketitle

\section{Introduction}

Deep inelastic scattering (DIS) processes enable precision studies of proton and nuclear structure thanks to the pointlike structure of the electron probe. The vast amount of accurate measurements in electron-proton collisions at HERA has allowed for a detailed determination of the partonic structure of the proton. In particular, a rapid increase of the gluon density towards small momentum fraction $x$ has been observed~\cite{Aaron:2009aa,Abramowicz:2015mha}. This  rise can not continue indefinitely without violating unitarity, and  nonlinear dynamics should eventually start to affect the proton or nuclear structure at small $x$. 

In order to describe the hadron structure at high densities where nonlinear dynamics should be included, an effective theory of quantum chromodynamics known as the Color Glass Condensate (CGC) has been developed, see Refs.~\cite{Iancu:2003xm,Gelis:2010nm,Blaizot:2016qgz} for a review. Despite the success of the CGC framework in describing high-energy scattering measurements,
there is no solid evidence that  nonlinear saturation effects are visible at current collider energies. In order to access more pronounced nonlinear effects, there are concrete plans to construct an Electron-Ion Collider in the US~\cite{Accardi:2012qut,Aschenauer:2017jsk,AbdulKhalek:2021gbh} with similar longer term plans at CERN~\cite{AbelleiraFernandez:2012cc,Agostini:2020fmq} and in China~\cite{Anderle:2021wcy}. As the parton densities are enhanced by approximatively $A^{1/3}$
in heavy nuclei, the nonlinear dynamics should be more easily accessible by replacing the proton by a heavy nucleus in these future facilities.

Exclusive vector meson production is an especially powerful tool in probing  hadron structure at small-$x$. In exclusive processes there cannot be a net color charge transfer in the process which, at leading order, requires an exchange of two gluons at the amplitude level. This renders the cross section to be approximatively sensitive to the squared gluon density~\cite{Ryskin:1992ui}, and nonlinear dynamics is expected to be pronounced in exclusive processes with heavy nuclear targets, see e.g. Ref.~\cite{Mantysaari:2017slo}. An additional benefit of  exclusive processes is the fact that only in these events it is possible to measure the total momentum transfer to the target, which is the Fourier conjugate to the impact parameter. This enables studies of the Generalized Parton Distribution Functions (GPDs)~\cite{Goeke:2001tz,Belitsky:2005qn} and the spatial structure of protons and nuclei with event-by-event fluctuations~\cite{Mantysaari:2020axf,Klein:2019qfb}.

So far, almost all phenomenological applications in the CGC framework have been at leading order in the QCD coupling constant $\as$, with the $\as \ln 1/x$ contributions resummed to all orders in terms of small-$x$ evolution equations such as the Balitsky-Kovchegov (BK) equation~\cite{Kovchegov:1999yj,Balitsky:1995ub}. Additionally, a subset of higher-order corrections to the evolution equation is included in terms of the
running coupling corrections~\cite{Balitsky:2006wa,Kovchegov:2006vj}. These applications include, for example, a successful description of the structure function~\cite{Albacete:2010sy,Lappi:2013zma} and vector meson production data~\cite{Kowalski:2006hc,Armesto:2014sma,Goncalves:2005yr,Cepila:2017nef,Lappi:2013am,Mantysaari:2018zdd,Mantysaari:2018nng,Mantysaari:2016jaz,Mantysaari:2016ykx,Mantysaari:2017dwh} measured at HERA~\cite{Chekanov:2002xi,Chekanov:2004mw,Alexa:2013xxa} and in Ultra Peripheral collisions~\cite{Bertulani:2005ru,Klein:2019qfb} at the LHC~\cite{Aaij:2014iea,Acharya:2018jua,Li:2020ntd,Khachatryan:2016qhq,Acharya:2019vlb}. In recent years, the theory has been developed towards next-to-leading order accuracy, including derivations of the next-to-leading order evolution equations~\cite{Balitsky:2008zza,Kovner:2013ona,Balitsky:2013fea,Lappi:2015fma,Lappi:2016fmu} with perturbative corrections to initial conditions~\cite{Dumitru:2020gla,Dumitru:2021tvw}, and impact factors for structure functions~\cite{Hanninen:2017ddy,Beuf:2016wdz,Lappi:2016oup,Ducloue:2017ftk,Beuf:2021qqa}, exclusive~\cite{Boussarie:2016bkq,Escobedo:2019bxn} and inclusive particle production~\cite{Ducloue:2017dit,Ducloue:2016shw,Stasto:2013cha,Chirilli:2011km,Chirilli:2012jd,Altinoluk:2014eka,Watanabe:2015tja,Iancu:2016vyg}.

In this Letter we present the first calculation of the exclusive heavy vector meson production at next-to-leading order (NLO) in photon-proton collisions. The advantage of heavy mesons such as the \jpsi is that the mass scale renders the process perturbative even at low virtualities $Q^2$, enabling  perturbative studies also in ultra peripheral collisions at RHIC and at the LHC where the photons are approximatively real. We include both the first QCD correction $\sim \as$, and the first nonrelativistic correction $\sim v^2$. We focus on \jpsi production, in which case both contributions can be expected to be parametrically important as for the charm quark velocity one can estimate $v^2 \sim \as$~\cite{Bodwin:2007fz}.
These developments are crucial to enable precision studies of nonlinear dynamics in exclusive scattering processes in the CGC framework. In this work we consider the case where the photon is longitudinally polarized, but note that it will be possible to extend the results to the transverse photon case in the future.

\section{Exclusive scattering at high energy}
\label{eq:exclusive}
At high energies, it is convenient to describe  exclusive vector meson production in the dipole picture in the frame where a highly energetic photon scatters off the target proton or nucleus and forms a vector meson. The exclusive nature of the process requires that there is no net color exchanged in the process. In this frame, the partonic Fock states of the virtual photon, e.g. $|q\bar q\rangle$ dipole and $|q\bar q g\rangle$ at NLO, have a long lifetime compared to the timescale of the interaction. These partons propagate eikonally through the color field of the target at fixed transverse coordinates $\xt_i$ and pick up Wilson lines $V(\xt_i)$ in the fundamental (quarks) or adjoint (gluons) representation.
The scattering amplitude can be conveniently written as a convolution of the photon and vector meson wave functions $\Psi_{\gamma^*}$ and $\Psi_V$, and Wilson line operators in the mixed longitudinal momentum fraction $z_i$, transverse coordinate $\xt_i$ space. At high energies the imaginary part dominates, and the scattering amplitude at NLO in the case where the transverse momentum transfer to the target vanishes can be written as
\begin{multline}
    \label{eq:im_amplitude}
    -i \acal = 2 \int_{\xt_0 \xt_1} \int \frac{\dd[]{z_0}\dd[]{z_1}}{(4\pi)} \delta(z_0+z_1-1) \Psi^{q \bar q*}_V \Psi_{\gamma^*}^{q \bar q} N_{01} \\
    + 2\int_{\xt_0 \xt_1 \xt_2} \int \frac{\dd[]{z_0}\dd[]{z_1}\dd[]{z_2}}{(4\pi)^2} \delta(z_0+z_1+z_2-1) \Psi^{q \bar q g*}_V \Psi_{\gamma^*}^{q \bar q g}  N_{012}.
\end{multline} 
Here $\xt_0,\xt_1$ and $\xt_2$ are the quark, antiquark and gluon transverse coordinates respectively, and $z_0, z_1$ and $z_2$ are the fractions of the photon plus momentum carried by these partons.
The calculations are done in the target rest frame where the photon plus momentum is chosen to be large. 
The wave functions depend implicitly on quark and gluon helicities, and a summation over the helicity states is also implicit.
We consider coherent scattering processes, where the target proton remains in the same quantum state (i.e. no target breakup). The coherent vector meson production cross section reads
\begin{equation}
    \label{eq:cross_section}
    \frac{\dd[]{}}{\dd[]{t}} \sigma^{\gamma^*+A \rightarrow V+A} = \frac{1}{16\pi} |\acal|^2.
\end{equation}
The Wilson line operator describing the $q\bar q$ dipole-target scattering known as the dipole amplitude $N_{01}$ is
\begin{equation}
    1-N_{01} = S_{01}=\Re \frac{1}{\nc} \left\langle V(\xt_0) V^\dagger(\xt_1) \right\rangle,
\end{equation}
where $\xt_0$ and $\xt_1$ are the quark and antiquark transverse coordinates.
Here the average $\langle \rangle$ refers to the average of the target color charge configurations.
Using the Fierz identity, the operator for the $q\bar q g$-target scattering can be written as (see e.g. Ref.~\cite{Hanninen:2017ddy})
\begin{equation}
	\label{eq:S-matrix_012}
	1-N_{012} = \frac{N_c}{2 C_F} \left( S_{02} S_{12} - \frac{1}{N_c^2} S_{01}\right).
\end{equation}
The transverse momentum transfer $|\Deltat| \approx \sqrt{-t}$ is the Fourier conjugate to the impact parameter. Consequently, an accurate calculation of the cross section differentially in  $|t|$  would require a realistic description of the impact parameter dependence in the small-$x$ evolution equation. Such equations exist and have been solved in the literature~\cite{Berger:2012wx,Berger:2010sh,Mantysaari:2018zdd,Bendova:2019psy}, but come with the price that one has to model confinement scale effects that suppress long range Coulomb tails. In this work, we want to perform a rigorous NLO calculation and limit our analysis to the $t=0$ case in which the diffractive scattering amplitude is only sensitive to the dipole-target scattering amplitude integrated over the impact parameter, which we take to satisfy an impact parameter independent small-$x$ evolution equation.

The necessary ingredients in the NLO calculation are the photon and vector meson wave functions and the dipole scattering amplitude, all at NLO level.
The wave functions have been recently calculated in the literature, first in the massless quark limit in Refs.~\cite{Hanninen:2017ddy,Beuf:2016wdz,Boussarie:2016bkq}, and recently the heavy quark contributions have become available~\cite{Beuf:2021qqa,Escobedo:2019bxn}. These wave functions are reviewed in Sec.~\ref{sec:nlo_vm}.
The dipole amplitude $N_{01}=N_{01}(Y)$ whose energy (or rapidity $Y$) dependence is given in terms of the  BK equation is also available at NLO accuracy. The BK equation requires a non-perturbative input that can be taken to describe the dipole-target scattering amplitude at initial $x$, typically around $x\sim 0.01$. This perturbative evolution then predicts the dipole amplitude at smaller $x$ (higher energies). In this work we use a BK evolved dipole scattering amplitude with the initial condition fitted to the HERA structure function data~\cite{Aaron:2009aa,Abramowicz:2015mha} at NLO accuracy in  Ref.~\cite{Beuf:2020dxl} (including only light quarks), using the public codes from Ref.~\cite{heikki_mantysaari_2020_4229269}.

\section{Vector meson production at next to leading order}
\label{sec:nlo_vm}
In order to calculate exclusive longitudinal quarkonium production at next-to-leading order accuracy, light front wave functions for the longitudinally polarized virtual photon $\Psi_{\gamma^*}$ and vector meson $\Psi_V$ are needed at this order in $\as$ (see also Ref.~\cite{Mantysaari:2020lhf} for a discussion of negligibly small polarization changing contributions). 

The virtual photon light front wave function at NLO accuracy in the case of massive quarks has recently been calculated in Ref.~\cite{Beuf:2021qqa}. In our NLO calculation, we need the wave function describing the photon fluctuation to the $|q\bar q\rangle$ Fock state at NLO, $\Psi_{\gamma^*}^{q\bar q}$, and the tree level result for the formation of the $|q\bar qg\rangle$ state, $\Psi_{\gamma^*}^{q\bar q g}$. Detailed expressions can be found from Ref.~\cite{Beuf:2021qqa}, and we also explicitly show these results in Appendix~\ref{appendix:photon}, equations \eqref{eq:photon_qq_NLO} and \eqref{eq:photon_qqg_NLO}.

For the heavy quarkonium, a systematic method to include both the higher order QCD corrections and the relativistic corrections has been derived in Ref.~\cite{Escobedo:2019bxn}. In this approach, the quarkonium wave function is written in terms of the coefficients $C^k_{n \leftarrow m}$ defined as 
\begin{equation}
	\label{eq:NR_expansion}
	\Psi^n_{V} = \sum_{m,k} C^k_{n \leftarrow m}
	 \int_0^1 \frac{\dd{z'}}{4\pi}  \left(\frac{1}{m_q}\nabla\right)^k\phi^m(\rt = 0,z'),
\end{equation}
where $\nabla = (\partial_{r_1},\partial_{r_2},(z'-1/2)2m_q i)$, $M_V$ and $m_q=M_V/2$ are the meson and heavy quark masses, and $\phi^{m}$  is the leading order light front wave function which is generally nonrelativistic in the case of heavy quarkonium. The transverse separation between the quark and the antiquark is $\rt$, and $z'$ is the fraction of the meson plus momentum carried by the quark.
Here $k=(k_1, k_2, k_3)$ is to be understood as a multi-index, and the sum goes over all positive values of the single indices $k_i$, and $\left(\frac{1}{m_q}\nabla\right)^k = \frac{1}{m_q^{|k|}} \nabla_1^{k_1} \nabla_2^{k_2} \nabla_3^{k_3}$.
As we neglect contributions $\sim \as v^2$, we need the quarkonium wave function $\Psi^n_V$ at NLO in the nonrelativistic limit, which corresponds to the $|k|=k_1+k_2+k_3=0$ case. The $|k|>0$
terms are suppressed by quark velocity, and are used to calculate relativistic corrections in Sec.~\ref{sec:relativistic}. The variables $m$ and $n$ refer to the partonic Fock states, and at  NLO in the nonrelativistic limit the non-zero coefficients are $C^{(0,0,0)}_{q\bar q\leftarrow q\bar q}$ and $C^{(0,0,0)}_{q\bar q g \leftarrow q\bar q}$. These have been calculated in Ref.~\cite{Escobedo:2019bxn} and are explicitly shown in Appendix~\ref{appendix:vm} in Eqs.~\eqref{eq:C0_qq} and \eqref{eq:C0_qqg}. 

With $m=q\bar q$, the leading order wave function for the longitudinally polarized quarkonium is $\phi^{q\bar q}_{h_0' h_1'}$ with the helicity structure given by $\delta_{h_0',-h_1'}$. Here $h_0'$ and $h_1'$ are the quark and antiquark helicities.  The coefficients $C^k_{n \leftarrow m}$ depend implicitly on the helicities of all the partons in state $n$, denoted by $h_i$, as well as on $h_0'$ and $h_1'$, and on parton colors as shown explicitly in Appendix~\ref{appendix:vm}. Summation over  helicities and colors is implicit in Eq.~\eqref{eq:NR_expansion}.
In particular, the 
 helicity structure of the $\Psi_V^{q\bar q}$ wave function becomes $\delta_{h_0,-h_1}$ at lowest order in $v$, where $h_0$ and $h_1$ are the quark and antiquark helicities. 
The helicity flip contribution (where the quark helicities are the same) would only appear in the longitudinal quarkonium wave function $\Psi_V^{q\bar q}$ at the order $v^4$~\cite{Lappi:2020ufv} which is not included in our calculation. Consequently we can also drop the helicity flip component $\Psi_{h.f}$ from the longitudinal photon wave function at NLO.

Using the NLO wave functions available in the literature, we calculate longitudinally polarized exclusive heavy vector meson production amplitude $-i \acal$ at next-to-leading order in the nonrelativistic limit. In this calculation there are both ultraviolet (UV) and infrared divergences that cancel at the level of the total NLO amplitude.
First, both the real $q\bar q g$ and virtual $q\bar q$ contributions contain ultraviolet divergences that cancel in the sum. In practice, we use dimensional regularization and subtract this UV divergence from the real contribution, and add it to the virtual contribution which renders both terms finite. As the UV subtraction term can contain an arbitrary finite piece,  division of the NLO contributions between the ``real'' and ``virtual'' parts is not unique. In this work, we follow the UV subtraction scheme developed in Refs.~\cite{Hanninen:2017ddy,Beuf:2021qqa}  which differs slightly from the one used in Ref.~\cite{Escobedo:2019bxn} and results in simpler expressions. 

In addition to UV divergences, the NLO amplitude is singular in the infrared where the gluon plus momentum is very small. In order to cancel this divergence, we include two contributions as discussed in Ref.~\cite{Escobedo:2019bxn}. First, the leading order wave function $\phi_{h_0' h_1'}^{q \bar q}$ is divergent at NLO, with the soft gluon divergence regulated by an infrared regulator $\alpha$. The leptonic decay width $\Gamma(V\to e^-e^+)$, however, is finite and connects the  wave function and the infrared regulator:~\cite{Escobedo:2019bxn} 
\begin{multline}
    \label{eq:Gamma_ee}
    \Gamma(V \rightarrow e^- e^+) = \frac{2N_c e_f^2 e^4}{3 \pi M_V}   \sum_{h_0' h_1'} \left|\int \frac{\dd[]{z'}}{4 \pi} \phi^{q \bar q}_{h_0'h_1'}\right|^2 \\
    \times \left[ 1 + \frac{2 \alpha_s C_F}{\pi} \left( \frac{1}{2\alpha}-2 \right) \right].
\end{multline}
Here $e$ is the elementary charge and $e_f$ the fractional charge of the quark.
In practice the wave function $\phi^{q\bar q}_{h_0'h_1'}$ can be written in terms of the decay width and the infrared regulator, and the $1/\alpha$ divergence will cancel when combined with the virtual NLO contribution. Additionally, the real gluon emission contribution is also singular in the soft gluon limit. This contribution can be absorbed in the target BK evolution.

The final result for the scattering amplitude at next-to-leading order reads
\begin{multline}
	\label{eq:total_NLO}
		-i \acal^L = -Q\sqrt{ \Gamma(V \rightarrow e^- e^+) \frac{3  M_V}{16\pi^2 \alpha_\textrm{em}} }  \int \dd[2]{\xt_{01}} \int \dd[2]{\bt} \\
		\Bigg\{ \kcal_{q \bar q}^\lo(Y_0) 	+\frac{\alpha_s C_F}{2\pi} \kcal_{q \bar q}^\nlo(\Ydip) \\ 
		+ \frac{\alpha_s C_F}{2\pi}\int \dd[2]{\xt_{20}} \int_{\zmin}^{1/2} \dd[]{z_2} \kcal_{q\bar q g}(Y_\text{qqg})\Bigg\}.
	\end{multline} 
	where $\kcal_{q \bar q}^\lo(Y_0) = K_0(\zeta) N_{01}(Y_0) $, $\zeta =|\xt_{01}|\sqrt{\frac{1}{4}Q^2+m_q^2} $, $\xt_{ij} = \xt_i -\xt_j$ and $\bt$ is the impact parameter. Detailed expressions for the NLO contributions $\kcal_{q\bar q}^\nlo$ and $\kcal_{q\bar q g}$ are shown in Appendix~\ref{appendix:nlo_xs}, Eqs.~\eqref{eq:K_qq_NLO} and~\eqref{eq:K_qqg}.
This corresponds to the ``unsubstracted scheme'' discussed e.g. in Refs.~\cite{Ducloue:2017ftk,Beuf:2020dxl}. Following the same terminology, we refer to the second term in Eq.~\eqref{eq:total_NLO} as the virtual ``dipole'' contribution (denoted by $\nlodip$ later), and the third term as the real contribution ($\nloqqg$). As discussed above, the division of the NLO corrections between these two terms is not unique. The dipole amplitudes are evaluated at evolution rapidities $Y_0$, $\Ydip$ and $\Yqqg$ that are discussed in detail below.

\subsection*{Evolution equations and rapidities}
The integral of $\kcal_{q \bar q g}$ over $z_2$ in Eq.~\eqref{eq:total_NLO} is singular in the limit $\zmin \rightarrow 0$. The singular part is related to the rapidity evolution of the dipole amplitude as can be seen by writing out the singularity explicitly:
\begin{multline}
    \label{eq:K_qqg_sing}
    \int \dd[2]{\xt_{20}} \int_{\zmin}^{1/2} \dd[]{z_2} \kcal_{q\bar q g}= \textrm{nonsingular term} \\+ K_0(\zeta) \int \dd[2]{\xt_{20}} \int_{\zmin}^{1/2} \dd[]{z_2} \frac{2}{\pi z_2} \frac{\xt_{01}^2}{\xt_{20}^2 \xt_{21}^2}  (S_{01}-S_{012}).
\end{multline}
We can recognize from this the leading-order Balitsky-Kovchegov equation in  integral form. It corresponds to the evolution over $\ln(\frac{1}{2\zmin})$ units of projectile rapidity $Y$, defined as $Y=\ln k^+/P^+$. We recall that we work in the frame where the incoming photon has a large plus momentum $q^+$ and the gluon plus momentum reads $k^+ = z_2 q^+$. The target plus momentum $P^+$ is obtained as $P^+ = Q_0^2/(2P^-)$, where the transverse momentum scale of the target is taken to be $Q_0^2=1\gev^2$ following~\cite{Beuf:2020dxl}. The photon-proton center of mass energy squared is $W^2= 2 q^+ P^-$.

When the singular part in Eq.~\eqref{eq:K_qqg_sing} is combined with the term $\kcal^\lo_{q\bar q}(Y_0)$, one obtains the leading order contribution but with the dipole amplitude evolved from rapidity $Y_0$ to rapidity 
\begin{equation}
    \Ydip = Y_0 + \ln\frac{1}{2\zmin}
\end{equation}
using the LO BK equation at fixed coupling. This evolution is part of the actual leading order contribution, as the BK evolution resums $\as \ln 1/x$ contributions, that at high energy are of the order $1$, to all orders.  In this work we use the dipole amplitudes obtained as a result of the NLO fit to HERA structure function data~\cite{Beuf:2020dxl} where $Y_0=0$, and we also use the same running coupling prescription as in Ref.~\cite{Beuf:2020dxl} in numerical analysis. 
We can write the leading order scattering amplitude as
\begin{multline}
\label{eq:total_LO}
    -i \acal^L_\mathrm{LO} = -Q\sqrt{ \Gamma(V \rightarrow e^- e^+) \frac{3  M_V}{16\pi^2 \alpha_\textrm{em}} } \\
    \times \int \dd[2]{\xt_{01}}\dd[2]{\bt} \kcal^\lo_{q\bar q}(\Ydip).
\end{multline}

We note that what actually is the leading order contribution is not unique. One could as well define it as the sum of the lowest order contribution evaluated at $Y_0$ and the singular part of $\nloqqg$. These two definitions match at fixed coupling if no higher order corrections were included in the BK evolution~\cite{Ducloue:2017ftk}. In the NLO DIS fit of Ref.~\cite{Beuf:2020dxl} applied in this work, modified versions of the BK equation that include the most important higher order corrections, in addition to the running coupling effects, were used and consequently Eq.~\eqref{eq:total_LO} 
can not be obtained from Eq.~\eqref{eq:total_NLO}. 
The definition of what is considered as the leading order contribution matters, because
when calculating the cross sections an interference between the leading order and the genuine next-to-leading order contributions is needed, in which case the NLO correction is obtained as $-i\acal^L - (-i)\acal^L_\mathrm{LO}$.

Let us then determine the lower limit of the $z_2$ integral $\zmin$ which controls the  amount of small-$x$ evolution. 
The applicability of the eikonal approximation requires that the invariant mass of the $q\bar q g$ system $M_{q\bar q g}$ satisfies  $M_{q\bar qg}^2 \ll W^2$. There is some freedom in determining how strong ordering is required, and resulting differences at the cross section level will again formally be of higher order in $\as$. In this work we use the same convention as in Ref.~\cite{Beuf:2020dxl} and require $M_{q\bar qg}^2 < W^2$, which gives 
\begin{equation}
\label{eq:z2min}
    z_2 > \zmin = \frac{P^+}{q^+}=  \frac{Q_0^2}{W^2 + Q^2 - m_N^2}.
\end{equation}
In the $\nloqqg$ term the rapidity at which the dipole amplitudes are evaluated depends on $z_2$, and using again $Y=z_2 q^+/P^+$ we get 
\begin{equation}
\label{eq:Yqqg}
    \Yqqg = \ln z_2 + \ln \frac{W^2 + Q^2 - m_N^2}{Q_0^2}.
\end{equation}
We note that the total evolution range probed in the $\nlodip$ contribution is exactly $\Ydip$ discussed above.

For consistency, we  choose to evaluate the dipole amplitude in the $\nlodip$ term at the same rapidity $\Ydip$ as the leading order contribution. In Ref.~\cite{Beuf:2020dxl} the virtual corrections to the structure functions were evaluated at $Y=\ln 1/\xbj$,  which in our case would correspond to the rapidity $Y_\mathrm{dip}^\mathrm{incl} = \ln 1/\xpom \neq \Ydip$, where $\xpom \approx (M_V^2+Q^2)/(W^2+Q^2)$  is the fraction of the target longitudinal momentum transferred  to the meson. Although it is not exactly consistent to use a different scheme to set the evolution rapidity in the structure function fit and in the application of these fit results in exclusive vector meson production, here we choose to apply the more natural choice for the evolution rapidity. The difference between these two choices is formally of higher order in $\as$.

In Ref.~\cite{Beuf:2020dxl} the fits are performed using initial conditions parametrized at both at $\Ybk=0$ and at $\Ybk=4.61$, in which case there is no evolution in the region $0<Y<\Ybk$.
The evolution equations that approximate the full NLO BK~\cite{Balitsky:2008zza} in the fits are the so-called KCBK, ResumBK and TBK equations (following the terminology of Ref.~\cite{Beuf:2020dxl}) derived in Refs.~\cite{Iancu:2015vea,Iancu:2015joa,Beuf:2014uia,Ducloue:2019ezk}. 
The ``kinematically constrained BK equation'' (KCBK)~\cite{Beuf:2014uia} is obtained by explicitly enforcing the required time ordering between the subsequent gluon emissions in the evolution. This procedure effectively resums corrections that are enhanced by two large transverse logarithms $\sim \as \ln \frac{\xt_{02}^2}{\xt_{01}^2} \ln \frac{\xt_{12}^2}{\xt_{02}^2} $ in the evolution, and the same double logarithms are also resummed in Ref.~\cite{Iancu:2015vea}. When additional contributions enhanced by single transverse logarithms $\sim \as \ln 1/(\xt_{ij}^2 \qs^2)$ (where $\qs$ is the saturation scale of the target) are also resummed following Ref.~\cite{Iancu:2015joa} one obtains the  evolution equation referred to as the ResumBK equation.
The third evolution equation (TBK) refers to the BK equation where the evolution rapidity $\eta$ (``target rapidity'')  is related to the fraction of the total longitudinal momentum of the target. When using the fit result that is written in terms of the target rapidity $\eta$ in the impact factors written in terms of the (projectile) rapidity $Y$, we apply the same shift as in Ref.~\cite{Beuf:2020dxl}: $\eta = Y+\ln(\min(1,\xt_{01}^2Q_0^2))$.
For more details of the different evolution equations, we refer the reader to Ref.~\cite{Beuf:2020dxl}.

\section{Relativistic corrections}
\label{sec:relativistic}
As we have parametrically $\alpha_s \sim v^2$, it is also interesting to consider the first relativistic corrections of order $v^2$ at leading order in $\alpha_s$. Using Eq.~\eqref{eq:NR_expansion}, we note that each term in the expansion corresponds to a correction of order $v^{|k|}$. The coefficient functions $C^k_{q\bar q \leftarrow q \bar q}$ are straightforward to calculate at leading order in $\as$ as then the wave function gets no loop corrections  (and $C^k_{q\bar q g \leftarrow q\bar q}=0$), and we can write
\begin{multline}
	\label{eq:NR_expansion_alphas0}
	\Psi^{q \bar q}_{V}(\alpha_s=0) = \sum_{k_1, k_2, k_3=0}^\infty C^{(k_1,k_2,k_3)}_{q\bar q \leftarrow q \bar q} \phi^{q \bar q}_{h_0' h_1'}(k_1,k_2,k_3),
\end{multline}
where 
\begin{multline}
	\label{eq:C-rel-constants}
   C^{(k_1,k_2,k_3)}_{q\bar q \leftarrow q \bar q} = \frac{\delta_{\alpha_0 \alpha_1}}{\sqrt{\nc}} \delta_{h_0 h_0'} \delta_{h_1 h_1'} \frac{1}{k_1!k_2!k_3!}  (m_q r_1)^{k_1} (m_q r_2)^{k_2} \\
	\times 4\pi\left(-\frac{1}{2i}\partial_z\right)^{k_3}\delta\left(z-1/2\right) \textrm{, and}
\end{multline}
\begin{multline}
	\label{eq:phi-constants}
    \phi^{q \bar q}_{h_0' h_1'}(k_1,k_2,k_3) \coloneqq \int_0^1 \frac{\dd{z'}}{4\pi} \frac{1}{m_q^{k_1+k_2}} \\
    \times \partial_{1}^{k_1} \partial_{2}^{k_2} \phi_{h_0' h_1'}^{q \bar q}(\rt = 0, z') [2i(z'-1/2)]^{k_3}.
\end{multline}
Here $\rt = (r_1,r_2)$ is the transverse separation of the two quarks and $\alpha_0,\alpha_1$ refer to the quark colors.

Calculating the production amplitude at order $v^2$ corresponds to keeping terms with $k_1+k_2+k_3 \leq 2$. The nonperturbative constants $ \phi^{q \bar q}_{h_0' h_1'}(k_1,k_2,k_3)$  can be related to the derivatives of the leading-order rest frame wave function $\phi_\text{RF}$ at the origin as shown in Ref.~\cite{Lappi:2020ufv}. This allows us to write (see discussion in Appendix~\ref{appendix:vm} for more details)
\begin{multline}
	\label{eq:v^2_LFWF}
		\phi^{q \bar q}_{h_0' h_1'}(2,0,0) = \phi^{q \bar q}_{h_0' h_1'}(0,2,0) =\phi^{q \bar q}_{h_0' h_1'}(0,0,2) \\
		= \frac{1}{\sqrt{2}} \delta_{h_0,-h_1}\frac{1}{\sqrt{m_q}} \frac{\nabla^2 \phi_\text{RF}(0)}{6m_q^2}.
\end{multline}
With this the order $v^2$ correction to the production amplitude is:
\begin{multline}
	\label{eq:total_relativistic2}
		-i \acal^L_\text{rel} = -\frac{e e_f Q \sqrt{N_c}}{2\pi\sqrt{2}}2 \int \dd[2]{\xt_{01}}  \int\dd[2]{\bt} N_{01}(\Ydip) \\
		\times \frac{\nabla^2 \phi_\text{RF}(0)}{12m_q^2\sqrt{m_q}} 
		 \left[ 2K_0(\zeta)-\frac{Q^2 \xt_{01}^2}{4\zeta}K_1(\zeta)+ m_q^2 \xt_{01}^2 K_0(\zeta) \right].
\end{multline}
The value for $\nabla^2 \phi_\text{RF}(0)$ for \jpsi can be determined using the nonrelativistic QCD (NRQCD) matrix elements from~\cite{Bodwin:2007fz}:
\begin{equation}
\label{eq:nabla^2 phi}
    \nabla^2 \phi_\text{RF}(0) = -\sqrt{\frac{\OLDME}{2N_c}} \qLDME = -0.120 \pm 0.039 \gev^{7/2}.
\end{equation}
The long-distance matrix elements $\OLDME$ and $\qLDME$ are related to the \jpsi wave function and its derivative at the origin and explicitly defined in Ref.~\cite{Bodwin:2007fz}. They are determined using $m_q=1.4\gev$ for the charm mass. In this work, on the other hand, we use the nonrelativistic limit relation $m_q=M_V/2$ also when calculating the relativistic correction. This difference is of higher order in $v$, see also discussion in Ref.~\cite{Lappi:2020ufv}.

\section{Numerical results}
In this section we present numerical results for the exclusive \jpsi production at $t=0$ in the kinematics covered by HERA and future EIC and LHeC/FCC-he measurements. Unless otherwise stated, we use the KCBK evolved dipole amplitude with the initial condition parametrized at $\Ybk=4.61$ from Ref.~\cite{Beuf:2020dxl}. The qualitative features do not depend on the actual dipole amplitude fit used.

\begin{figure}
    \centering
    \includegraphics[width=\columnwidth]{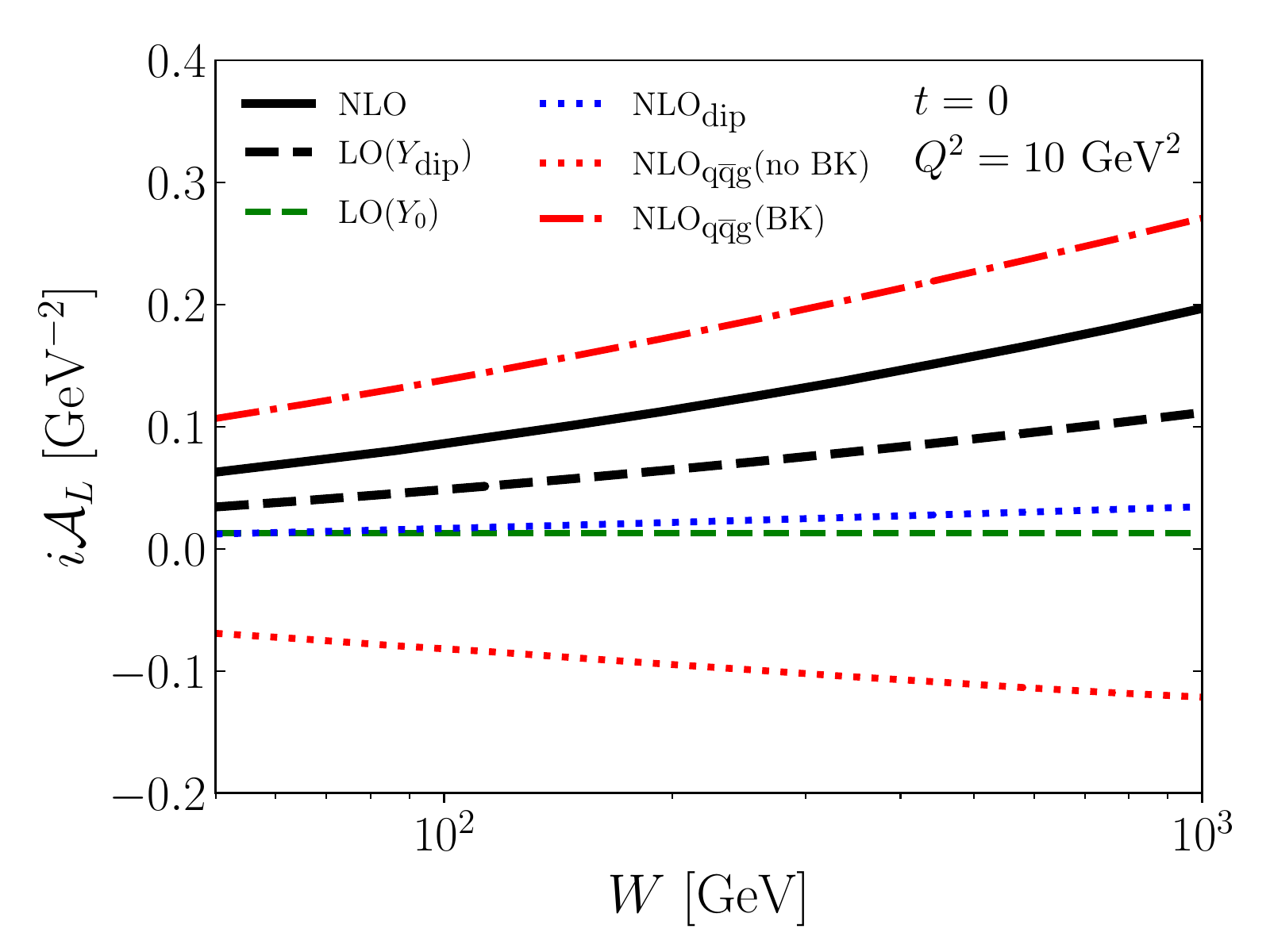}
    \caption{Different contributions to the exclusive \jpsi production scattering amplitude as a function of center-of-mass energy $W$.}
    \label{fig:amplitude_kcbk_wdep} 
\end{figure}

\begin{figure}
    \centering
    \includegraphics[width=\columnwidth]{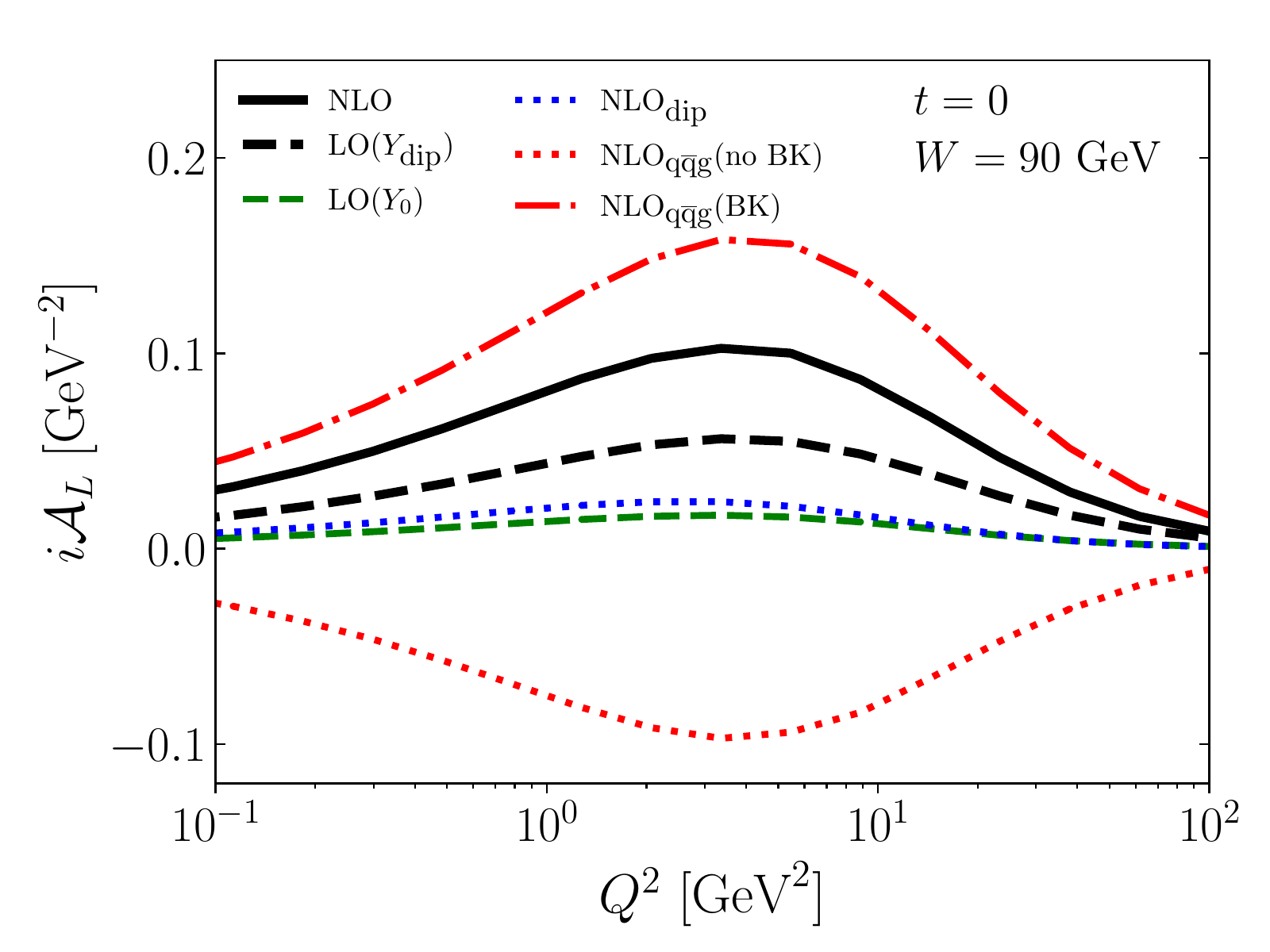}
    \caption{Different contributions to the exclusive \jpsi production scattering amplitude as a function of photon virtuality $Q^2$.}
    \label{fig:amplitude_kcbk_Q2dep}
\end{figure}

The scattering amplitudes for exclusive \jpsi production at leading and next-to-leading order 
 are shown in \figs\ref{fig:amplitude_kcbk_wdep} and \ref{fig:amplitude_kcbk_Q2dep} as a function of center-of-mass energy $W$ and photon virtuality $Q^2$.
The total NLO amplitude is shown in Eq.~\eqref{eq:total_NLO}, and should be compared to the leading-order result including the small-$x$ BK evolution defined in Eq.~\eqref{eq:total_LO} and denoted by LO$(\Ydip)$ in the figures. Note that all results in \figs\ref{fig:amplitude_kcbk_wdep} and \ref{fig:amplitude_kcbk_Q2dep} are obtained by using the same dipole amplitude $N_{01}$ from Ref.~\cite{Beuf:2020dxl}.
The NLO corrections are found to be sizeable, of the order of $\sim 75\%$, and depend weakly on $W$ and $Q^2$. Only at highest $Q^2$ values $\sim 100\gev^2$ (where the high scale renders $\as$ smaller) the NLO corrections become slightly less important.

In Figs.~\ref{fig:amplitude_kcbk_wdep} and \ref{fig:amplitude_kcbk_Q2dep} the different contributions to the NLO amplitude are also shown separately. First, the LO($Y_0)$ curve refers to the leading order result with no BK evolution. The virtual NLO correction $\nlodip$ is found to be small and positive (by positive we mean that it has the same sign as the leading order result) at all $W$ and $Q^2$. 
The real contribution $\nloqqg$ 
includes a leading order part in terms of the BK evolution.
In the figures we show separately the contribution from the BK evolution shown in Eq.~\eqref{eq:K_qqg_sing}, and 
the genuine next-to-leading order correction to it due to the exact gluon emission kinematics included in the full NLO calculation. This NLO correction $\nloqqg(\text{no BK}) = \nloqqg - \nloqqg(\mathrm{BK})$ significantly suppresses the effect of the small-$x$ BK evolution as expected.
This systematics in the real and virtual corrections is similar to what is observed in case of structure function calculations at NLO in Ref.~\cite{Ducloue:2017ftk}. However, we emphasize that the division of the NLO corrections between the $\nlodip$ and $\nloqqg$ terms is not unique, see the discussion in Sec.~\ref{sec:nlo_vm}.

\begin{figure}
    \centering
    \includegraphics[width=\columnwidth]{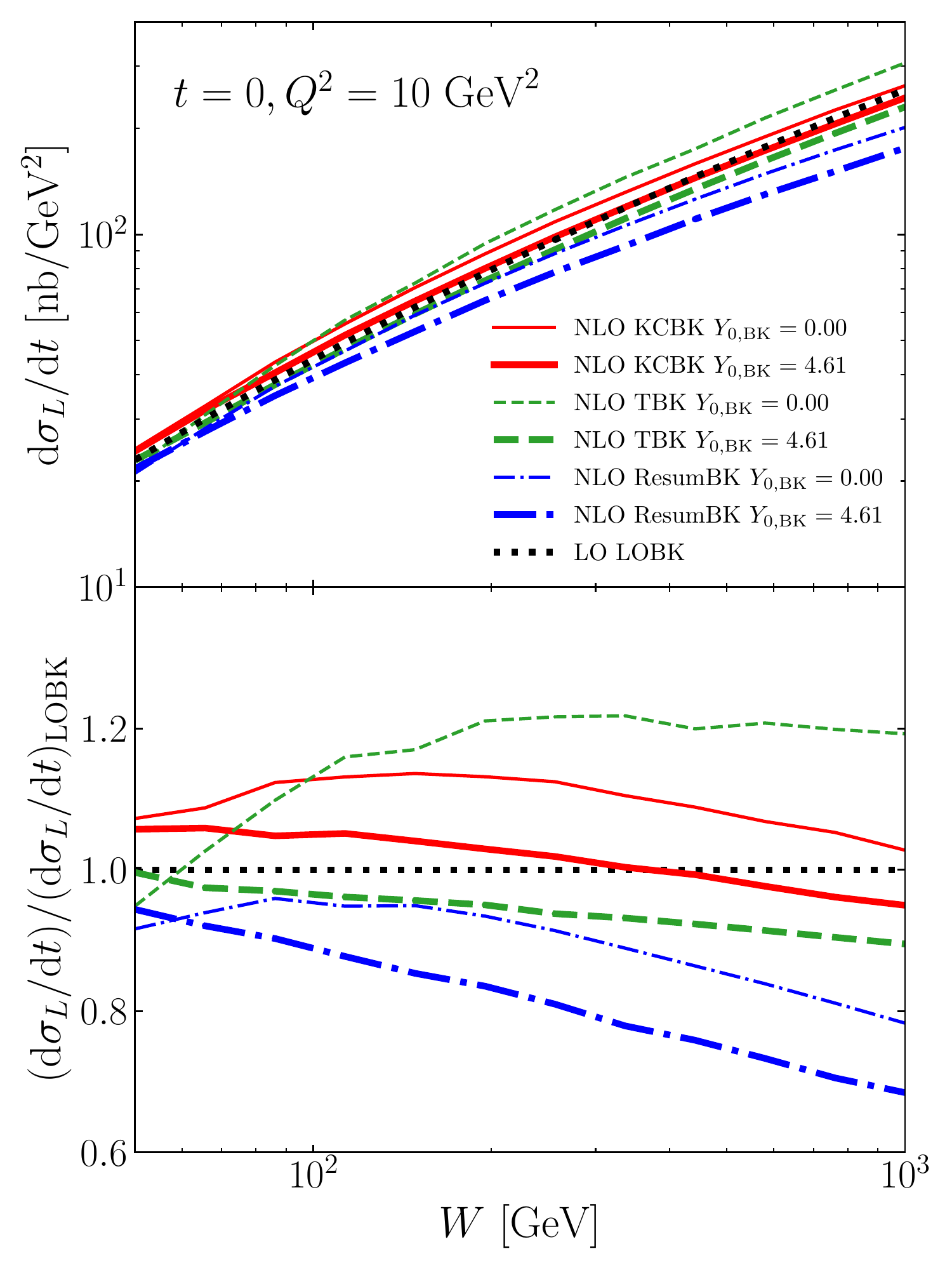}
    \caption{Center-of-mass energy dependence of the exclusive \jpsi electroproduction cross section at NLO. }
    \label{fig:xs_wdep}
\end{figure}

In \fig\ref{fig:xs_wdep} we show the energy dependence of the \jpsi electroproduction cross section at NLO using different dipole amplitude fits which describe the HERA structure function data approximatively equally well~\cite{Beuf:2020dxl}. For comparison, the LO result using the fit with LO BK evolution and leading order impact factors from Ref.~\cite{Lappi:2013zma} is also shown. In the case of the LO BK evolved result the evolution rapidity is chosen as $Y=\ln 0.01/\xpom$, consistently with the leading-order fit procedure of Ref.~\cite{Lappi:2013zma}. 

Despite the fact that all dipole amplitudes result in almost identical descriptions of the HERA structure function data, we find that the resulting \jpsi production cross sections can differ by almost a factor of $2$. This demonstrates that  vector meson production provides complementary information for the extraction of the initial condition for the BK evolved dipole scattering amplitude, 
as it is sensitive to the dipole-target interaction 
at different distance scales (see also Ref.~\cite{Mantysaari:2018nng,Mantysaari:2018zdd}) compared to total cross section measurements. 

It should be noted that the NLO results for the \jpsi production cross section are closer to the LO BK result than one might expect judging from \figs\ref{fig:amplitude_kcbk_wdep} and \ref{fig:amplitude_kcbk_Q2dep}. This can be understood by noting that the LO BK result in Fig.~\ref{fig:xs_wdep} is calculated using a LO dipole amplitude resulting from a leading-order fit, whereas in Figs.~\ref{fig:amplitude_kcbk_wdep} and~\ref{fig:amplitude_kcbk_Q2dep} the same NLO fit result was used for the dipole amplitude in all cases. In particular, the fit parameters obtained from the LO fit include an effective description of some of the higher-order effects. However, we still find that the NLO cross section generically evolves more slowly in $W$  at high energies compared to LO results.

\begin{figure}
    \centering
    \includegraphics[width=\columnwidth]{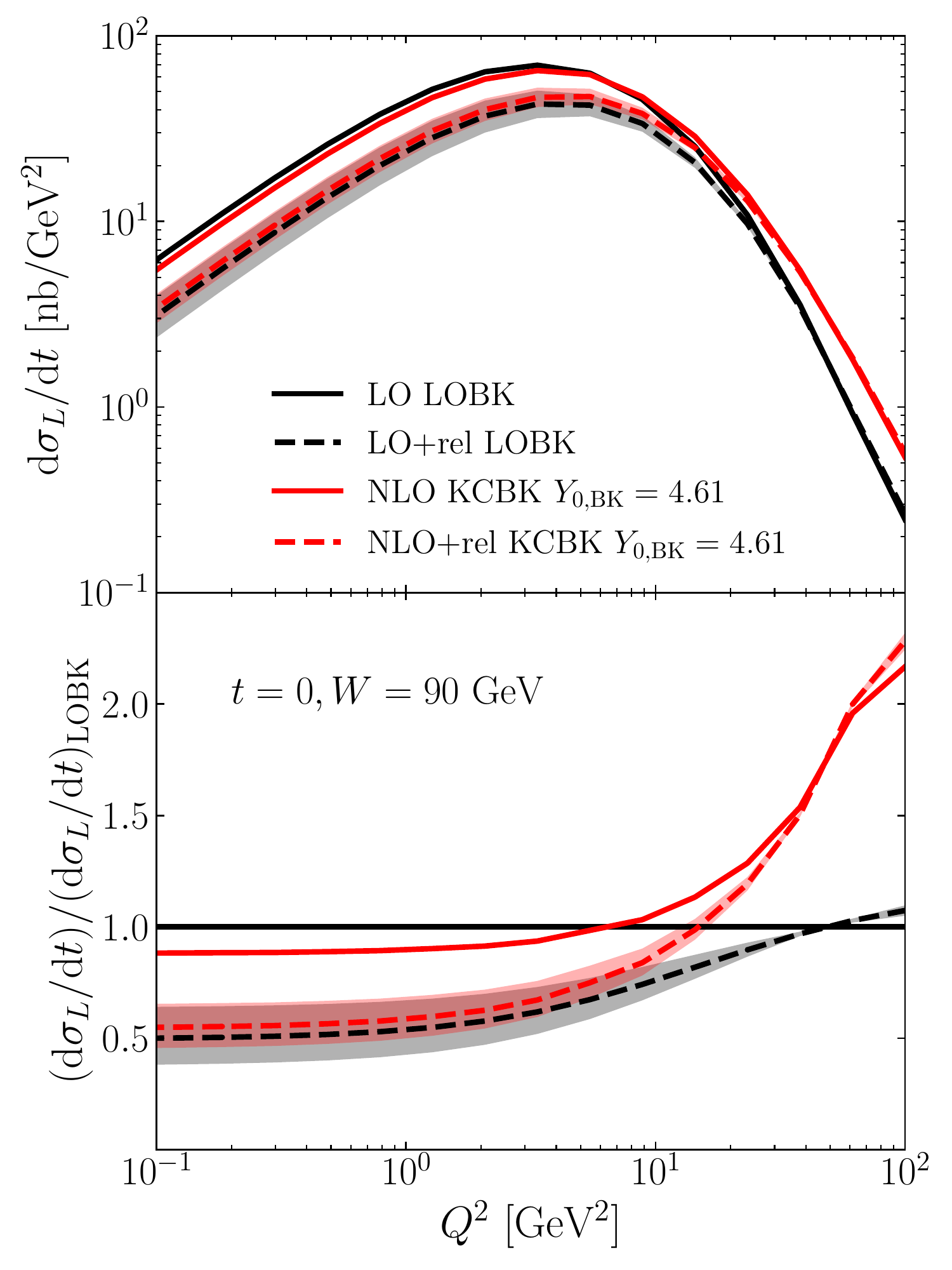}
    \caption{The effect of the relativistic correction on \jpsi production at LO and at NLO as a function of photon virtuality $Q^2$. The bands show the uncertainties of the relativistic corrections. }
    
    \label{fig:xs_qdep}
\end{figure}

In \fig\ref{fig:xs_qdep} the effect of the relativistic corrections is shown. As previously discussed in Ref.~\cite{Lappi:2020ufv}, the relativistic corrections are significant and decrease the cross section up to $\sim 50\%$ at low photon virtualities, and become insignificant (but non-zero~\cite{Hoodbhoy:1996zg}) at large $Q^2$. At low virtualities the relativistic correction is more important than the next-to-leading order contribution. However, when comparing the relativistic $\sim v^2$ and NLO $\sim \as$ corrections one has to keep in mind that the leading order BK evolution effectively includes higher order corrections encoded in the fit parameters as discussed above.
The relativistic correction is less important when it is added on top of the next-to-leading order result, $\sim 40 \%$ at low $Q^2$, as we do not include corrections of the order $\as v^2$.

The next-to-leading order correction becomes large at high virtualities as can be seen from the lower panel of Fig.~\ref{fig:xs_qdep}. We note that both LO and NLO fits provide a good description of the $Q^2$ dependence of the HERA structure function data at small $x$. The stronger virtuality dependence at leading order can be again understood to result from the fact that \jpsi production is sensitive to dipole scattering amplitude at smaller length scales compared to structure functions. The small dipole size region is also only weakly constrained by the structure function data when the initial condition for the BK evolution is fitted.  

Technically, the dependence on the virtuality is related to the anomalous dimension $\gamma$ which describes the behaviour of the dipole amplitude at small dipole sizes: $N_{01}\sim (\xt_{01}^2\qs^2)^\gamma$. At leading order the BK evolution results in $\gamma\sim 0.7$ at large rapidities, but as $Y\sim \ln 1/\xpom$ in the LO fit, at high $Q^2$ one is actually sensitive to the dipole amplitude close to the initial condition where $\gamma\sim 1.2$~\cite{Albacete:2010sy,Lappi:2013zma}. On the other hand, in our NLO setup there is a long evolution at high $Q^2$, see Eq.~\eqref{eq:Yqqg}. However, the anomalous dimension at asymptotically small dipoles does not actually change when higher order corrections are resummed in the NLO fit. As the NLO fits also result in $\gamma \sim 1.2$~\cite{Beuf:2020dxl} at the initial rapidity, in principle we would expect to see comparable $Q^2$ evolution speeds in the exclusive \jpsi production. In practice one is not probing the dipole amplitude at asymptotically small dipoles but at $\xt_{01}^2 \sim 1/Q^2$, and in the NLO fits $\gamma$ decreases in the evolution at intermediate dipole sizes~\cite{Beuf:2020dxl}. As a result, one finds that the NLO exclusive vector meson cross section decreases more slowly as a function of $Q^2$ than the leading order case at high virtualities.

\section{Conclusions}
We have calculated, for the first time, exclusive heavy vector meson production at next-to-leading order in the Color Glass Condensate framework. In the calculation we apply the recently derived wave functions for the virtual photon and  vector meson including massive quarks. The main result of this work, the scattering amplitude for  longitudinal vector meson production at NLO, is Eq.~\eqref{eq:total_NLO}. We emphasize that this result is free from any ultraviolet or infrared divergences and suitable for phenomenological applications.

We have numerically evaluated the derived scattering amplitude, using dipole-proton scattering amplitudes recently obtained as a result of an NLO fit to HERA structure function data. We have presented the first numerical calculation of the exclusive \jpsi production cross section at NLO in the CGC framework. As the future Electron-Ion Collider and other nuclear DIS facilities will provide vast amounts of precise vector meson production data in the future, these developments that promote the CGC calculations to the precision level are extremely important.

We have shown that the next-to-leading order corrections to the \jpsi production cross section are significant, although these corrections can partially be captured in leading order calculations by the non-perturbative fit parameters. We also demonstrate that the vector meson production data provides complementary information compared to structure function measurements. A global analysis including both the reduced cross section and exclusive vector meson production data would be preferable in the future when extracting the initial condition for the Balitsky-Kovchegov evolution of the dipole scattering amplitude.
Comparing the NLO $\sim \as$ correction to the relativistic $\sim v^2$ correction, we have observed that especially at low virtualities both corrections are numerically important, with the relativistic correction generically larger. 

In the future, we  will include the contribution from the transversely polarized virtual photons. This development will enable comparisons with the vector meson production data from HERA~\cite{Alexa:2013xxa,Chekanov:2002xi,Chekanov:2004mw} and from the UPC physics program at the LHC~\cite{Aaij:2014iea,Acharya:2018jua}, as well as calculation of precise predictions for the EIC. Extending the calculation from protons to heavy nuclei will also enable precision studies of saturation phenomena in current~\cite{Khachatryan:2016qhq,Acharya:2019vlb,Li:2020ntd} and future nuclear DIS experiments.

\section*{Acknowledgements}
We thank M. Escobedo, T. Lappi and R. Paatelainen for useful discussions and are grateful to authors of Ref.~\cite{Beuf:2021qqa} for sharing their results before publication.
This work was supported by the Academy of Finland, projects 314764, 338263 and 346567 (H.M) and 321840 (J.P),  and by the EU Horizon 2020 research and innovation programme, STRONG-2020 project (Grant Agreement No. 824093). The content of this article does not reflect the official opinion of the European Union and responsibility for the information and views expressed therein lies entirely with the authors.

\pagebreak

\bibliographystyle{JHEP-2modlong.bst}
\bibliography{refs}

\providecommand{\href}[2]{#2}\begingroup\raggedright\begin{thebibliography}{10}

\bibitem{Aaron:2009aa}
{\bf H1, ZEUS} collaborations, F.~D. Aaron {\em et.~al.}, {\it {Combined
  Measurement and QCD Analysis of the Inclusive $e^\pm p$ Scattering Cross
  Sections at HERA}},  \href{http://dx.doi.org/10.1007/JHEP01(2010)109}{{\em
  JHEP} {\bf 01} (2010) 109} [\href{http://arXiv.org/abs/0911.0884}{{\tt
  arXiv:0911.0884 [hep-ex]}}].

\bibitem{Abramowicz:2015mha}
{\bf H1, ZEUS} collaborations, H.~Abramowicz {\em et.~al.}, {\it {Combination of
  measurements of inclusive deep inelastic ${e^{\pm }p}$ scattering cross
  sections and QCD analysis of HERA data}},
  \href{http://dx.doi.org/10.1140/epjc/s10052-015-3710-4}{{\em Eur. Phys. J. C}
  {\bf 75} (2015)~no.~12 580} [\href{http://arXiv.org/abs/1506.06042}{{\tt
  arXiv:1506.06042 [hep-ex]}}].

\bibitem{Iancu:2003xm}
E.~Iancu and R.~Venugopalan, {\em {The Color glass condensate and high-energy
  scattering in QCD}}, pp.~249--3363.
\newblock 3, 2003.
\newblock \href{http://arXiv.org/abs/hep-ph/0303204}{{\tt
  arXiv:hep-ph/0303204}}.

\bibitem{Gelis:2010nm}
F.~Gelis, E.~Iancu, J.~Jalilian-Marian and R.~Venugopalan, {\it {The Color
  Glass Condensate}},
  \href{http://dx.doi.org/10.1146/annurev.nucl.010909.083629}{{\em Ann. Rev.
  Nucl. Part. Sci.} {\bf 60} (2010) 463}
  [\href{http://arXiv.org/abs/1002.0333}{{\tt arXiv:1002.0333 [hep-ph]}}].

\bibitem{Blaizot:2016qgz}
J.-P. Blaizot, {\it {High gluon densities in heavy ion collisions}},
  \href{http://dx.doi.org/10.1088/1361-6633/aa5435}{{\em Rept. Prog. Phys.}
  {\bf 80} (2017)~no.~3 032301} [\href{http://arXiv.org/abs/1607.04448}{{\tt
  arXiv:1607.04448 [hep-ph]}}].

\bibitem{Accardi:2012qut}
A.~Accardi {\em et.~al.}, {\it {Electron Ion Collider: The Next QCD Frontier}:
  {Understanding the glue that binds us all}},
  \href{http://dx.doi.org/10.1140/epja/i2016-16268-9}{{\em Eur. Phys. J. A}
  {\bf 52} (2016)~no.~9 268} [\href{http://arXiv.org/abs/1212.1701}{{\tt
  arXiv:1212.1701 [nucl-ex]}}].

\bibitem{Aschenauer:2017jsk}
E.~C. Aschenauer, S.~Fazio, J.~H. Lee, H.~Mäntysaari, B.~S. Page, B.~Schenke,
  T.~Ullrich, R.~Venugopalan and P.~Zurita, {\it {The electron\textendash{}ion
  collider: assessing the energy dependence of key measurements}},
  \href{http://dx.doi.org/10.1088/1361-6633/aaf216}{{\em Rept. Prog. Phys.}
  {\bf 82} (2019)~no.~2 024301} [\href{http://arXiv.org/abs/1708.01527}{{\tt
  arXiv:1708.01527 [nucl-ex]}}].

\bibitem{AbdulKhalek:2021gbh}
R.~Abdul~Khalek {\em et.~al.}, {\it {Science Requirements and Detector Concepts
  for the Electron-Ion Collider: EIC Yellow Report}},
  \href{http://arXiv.org/abs/2103.05419}{{\tt arXiv:2103.05419
  [physics.ins-det]}}.

\bibitem{AbelleiraFernandez:2012cc}
{\bf LHeC Study Group} collaboration, J.~L. Abelleira~Fernandez {\em et.~al.},
  {\it {A Large Hadron Electron Collider at CERN: Report on the Physics and
  Design Concepts for Machine and Detector}},
  \href{http://dx.doi.org/10.1088/0954-3899/39/7/075001}{{\em J. Phys. G} {\bf
  39} (2012) 075001} [\href{http://arXiv.org/abs/1206.2913}{{\tt
  arXiv:1206.2913 [physics.acc-ph]}}].

\bibitem{Agostini:2020fmq}
{\bf LHeC and FCC-he Study Group} collaborations, P.~Agostini {\em et.~al.},
  {\it {The Large Hadron-Electron Collider at the HL-LHC}},
  \href{http://arXiv.org/abs/2007.14491}{{\tt arXiv:2007.14491 [hep-ex]}}.

\bibitem{Anderle:2021wcy}
D.~P. Anderle {\em et.~al.}, {\it {Electron-ion collider in China}},
  \href{http://dx.doi.org/10.1007/s11467-021-1062-0}{{\em Front. Phys.
  (Beijing)} {\bf 16} (2021)~no.~6 64701}
  [\href{http://arXiv.org/abs/2102.09222}{{\tt arXiv:2102.09222 [nucl-ex]}}].

\bibitem{Ryskin:1992ui}
M.~Ryskin, {\it {Diffractive $J/\Psi$ electroproduction in LLA QCD}},
  \href{http://dx.doi.org/10.1007/BF01555742}{{\em Z. Phys.} {\bf C57} (1993)
  89}.
%%CITATION = ZEPYA,C57,89;%%

\bibitem{Mantysaari:2017slo}
H.~M\"antysaari and R.~Venugopalan, {\it {Systematics of strong nuclear
  amplification of gluon saturation from exclusive vector meson production in
  high energy electron\textendash{}nucleus collisions}},
  \href{http://dx.doi.org/10.1016/j.physletb.2018.04.044}{{\em Phys. Lett. B}
  {\bf 781} (2018) 664} [\href{http://arXiv.org/abs/1712.02508}{{\tt
  arXiv:1712.02508 [nucl-th]}}].

\bibitem{Goeke:2001tz}
K.~Goeke, M.~V. Polyakov and M.~Vanderhaeghen, {\it {Hard exclusive reactions
  and the structure of hadrons}},
  \href{http://dx.doi.org/10.1016/S0146-6410(01)00158-2}{{\em Prog. Part. Nucl.
  Phys.} {\bf 47} (2001) 401} [\href{http://arXiv.org/abs/hep-ph/0106012}{{\tt
  arXiv:hep-ph/0106012}}].

\bibitem{Belitsky:2005qn}
A.~V. Belitsky and A.~V. Radyushkin, {\it {Unraveling hadron structure with
  generalized parton distributions}},
  \href{http://dx.doi.org/10.1016/j.physrep.2005.06.002}{{\em Phys. Rept.} {\bf
  418} (2005) 1} [\href{http://arXiv.org/abs/hep-ph/0504030}{{\tt
  arXiv:hep-ph/0504030}}].

\bibitem{Mantysaari:2020axf}
H.~M\"antysaari, {\it {Review of proton and nuclear shape fluctuations at high
  energy}},  \href{http://dx.doi.org/10.1088/1361-6633/aba347}{{\em Rept. Prog.
  Phys.} {\bf 83} (2020)~no.~8 082201}
  [\href{http://arXiv.org/abs/2001.10705}{{\tt arXiv:2001.10705 [hep-ph]}}].

\bibitem{Klein:2019qfb}
S.~R. Klein and H.~M\"antysaari, {\it {Imaging the nucleus with high-energy
  photons}},  \href{http://dx.doi.org/10.1038/s42254-019-0107-6}{{\em Nature
  Rev. Phys.} {\bf 1} (2019)~no.~11 662}
  [\href{http://arXiv.org/abs/1910.10858}{{\tt arXiv:1910.10858 [hep-ex]}}].

\bibitem{Kovchegov:1999yj}
Y.~V. Kovchegov, {\it {Small-$x$ $F_2$ structure function of a nucleus
  including multiple pomeron exchanges}},
  \href{http://dx.doi.org/10.1103/PhysRevD.60.034008}{{\em Phys. Rev.} {\bf
  D60} (1999) 034008} [\href{http://arXiv.org/abs/hep-ph/9901281}{{\tt
  arXiv:hep-ph/9901281 [hep-ph]}}].

\bibitem{Balitsky:1995ub}
I.~Balitsky, {\it {Operator expansion for high-energy scattering}},
  \href{http://dx.doi.org/10.1016/0550-3213(95)00638-9}{{\em Nucl. Phys.} {\bf
  B463} (1996) 99} [\href{http://arXiv.org/abs/hep-ph/9509348}{{\tt
  arXiv:hep-ph/9509348}}].
%%CITATION = HEP-PH/9509348;%%

\bibitem{Balitsky:2006wa}
I.~Balitsky, {\it {Quark contribution to the small-x evolution of color
  dipole}},  \href{http://dx.doi.org/10.1103/PhysRevD.75.014001}{{\em Phys.
  Rev. D} {\bf 75} (2007) 014001}
  [\href{http://arXiv.org/abs/hep-ph/0609105}{{\tt arXiv:hep-ph/0609105}}].

\bibitem{Kovchegov:2006vj}
Y.~V. Kovchegov and H.~Weigert, {\it {Triumvirate of Running Couplings in
  Small-$x$ Evolution}},
  \href{http://dx.doi.org/10.1016/j.nuclphysa.2006.10.075}{{\em Nucl. Phys. A}
  {\bf 784} (2007) 188} [\href{http://arXiv.org/abs/hep-ph/0609090}{{\tt
  arXiv:hep-ph/0609090}}].

\bibitem{Albacete:2010sy}
J.~L. Albacete, N.~Armesto, J.~G. Milhano, P.~Quiroga-Arias and C.~A. Salgado,
  {\it {AAMQS: A non-linear QCD analysis of new HERA data at small-x including
  heavy quarks}},  \href{http://dx.doi.org/10.1140/epjc/s10052-011-1705-3}{{\em
  Eur. Phys. J. C} {\bf 71} (2011) 1705}
  [\href{http://arXiv.org/abs/1012.4408}{{\tt arXiv:1012.4408 [hep-ph]}}].

\bibitem{Lappi:2013zma}
T.~Lappi and H.~M\"antysaari, {\it {Single inclusive particle production at
  high energy from HERA data to proton-nucleus collisions}},
  \href{http://dx.doi.org/10.1103/PhysRevD.88.114020}{{\em Phys. Rev. D} {\bf
  88} (2013) 114020} [\href{http://arXiv.org/abs/1309.6963}{{\tt
  arXiv:1309.6963 [hep-ph]}}].

\bibitem{Kowalski:2006hc}
H.~Kowalski, L.~Motyka and G.~Watt, {\it {Exclusive diffractive processes at
  HERA within the dipole picture}},
  \href{http://dx.doi.org/10.1103/PhysRevD.74.074016}{{\em Phys. Rev. D} {\bf
  74} (2006) 074016} [\href{http://arXiv.org/abs/hep-ph/0606272}{{\tt
  arXiv:hep-ph/0606272}}].

\bibitem{Armesto:2014sma}
N.~Armesto and A.~H. Rezaeian, {\it {Exclusive vector meson production at high
  energies and gluon saturation}},
  \href{http://dx.doi.org/10.1103/PhysRevD.90.054003}{{\em Phys. Rev. D} {\bf
  90} (2014)~no.~5 054003} [\href{http://arXiv.org/abs/1402.4831}{{\tt
  arXiv:1402.4831 [hep-ph]}}].

\bibitem{Goncalves:2005yr}
V.~P. Goncalves and M.~V.~T. Machado, {\it {The QCD pomeron in ultraperipheral
  heavy ion collisions. IV. Photonuclear production of vector mesons}},
  \href{http://dx.doi.org/10.1140/epjc/s2005-02175-3}{{\em Eur. Phys. J. C}
  {\bf 40} (2005) 519} [\href{http://arXiv.org/abs/hep-ph/0501099}{{\tt
  arXiv:hep-ph/0501099}}].

\bibitem{Cepila:2017nef}
J.~Cepila, J.~G. Contreras and M.~Krelina, {\it {Coherent and incoherent
  $\mathrm{J/}\psi$ photonuclear production in an energy-dependent hot-spot
  model}},  \href{http://dx.doi.org/10.1103/PhysRevC.97.024901}{{\em Phys. Rev.
  C} {\bf 97} (2018)~no.~2 024901} [\href{http://arXiv.org/abs/1711.01855}{{\tt
  arXiv:1711.01855 [hep-ph]}}].

\bibitem{Lappi:2013am}
T.~Lappi and H.~Mäntysaari, {\it {$J/\psi$ production in ultraperipheral Pb+Pb
  and $p$+Pb collisions at energies available at the CERN Large Hadron
  Collider}},  \href{http://dx.doi.org/10.1103/PhysRevC.87.032201}{{\em Phys.
  Rev. C} {\bf 87} (2013)~no.~3 032201}
  [\href{http://arXiv.org/abs/1301.4095}{{\tt arXiv:1301.4095 [hep-ph]}}].

\bibitem{Mantysaari:2018zdd}
H.~M\"antysaari and B.~Schenke, {\it {Confronting impact parameter dependent
  JIMWLK evolution with HERA data}},
  \href{http://dx.doi.org/10.1103/PhysRevD.98.034013}{{\em Phys. Rev. D} {\bf
  98} (2018)~no.~3 034013} [\href{http://arXiv.org/abs/1806.06783}{{\tt
  arXiv:1806.06783 [hep-ph]}}].

\bibitem{Mantysaari:2018nng}
H.~M\"antysaari and P.~Zurita, {\it {In depth analysis of the combined HERA
  data in the dipole models with and without saturation}},
  \href{http://dx.doi.org/10.1103/PhysRevD.98.036002}{{\em Phys. Rev. D} {\bf
  98} (2018) 036002} [\href{http://arXiv.org/abs/1804.05311}{{\tt
  arXiv:1804.05311 [hep-ph]}}].

\bibitem{Mantysaari:2016jaz}
H.~M\"antysaari and B.~Schenke, {\it {Revealing proton shape fluctuations with
  incoherent diffraction at high energy}},
  \href{http://dx.doi.org/10.1103/PhysRevD.94.034042}{{\em Phys. Rev. D} {\bf
  94} (2016)~no.~3 034042} [\href{http://arXiv.org/abs/1607.01711}{{\tt
  arXiv:1607.01711 [hep-ph]}}].

\bibitem{Mantysaari:2016ykx}
H.~M\"antysaari and B.~Schenke, {\it {Evidence of strong proton shape
  fluctuations from incoherent diffraction}},
  \href{http://dx.doi.org/10.1103/PhysRevLett.117.052301}{{\em Phys. Rev.
  Lett.} {\bf 117} (2016)~no.~5 052301}
  [\href{http://arXiv.org/abs/1603.04349}{{\tt arXiv:1603.04349 [hep-ph]}}].

\bibitem{Mantysaari:2017dwh}
H.~M\"antysaari and B.~Schenke, {\it {Probing subnucleon scale fluctuations in
  ultraperipheral heavy ion collisions}},
  \href{http://dx.doi.org/10.1016/j.physletb.2017.07.063}{{\em Phys. Lett. B}
  {\bf 772} (2017) 832} [\href{http://arXiv.org/abs/1703.09256}{{\tt
  arXiv:1703.09256 [hep-ph]}}].

\bibitem{Chekanov:2002xi}
{\bf ZEUS} collaboration, S.~Chekanov {\em et.~al.}, {\it {Exclusive
  photoproduction of $J/\psi$ mesons at HERA}},
  \href{http://dx.doi.org/10.1007/s10052-002-0953-7}{{\em Eur. Phys. J. C} {\bf
  24} (2002) 345} [\href{http://arXiv.org/abs/hep-ex/0201043}{{\tt
  arXiv:hep-ex/0201043}}].

\bibitem{Chekanov:2004mw}
{\bf ZEUS} collaboration, S.~Chekanov {\em et.~al.}, {\it {Exclusive
  electroproduction of $\mathrm{J}/\psi$ mesons at HERA}},
  \href{http://dx.doi.org/10.1016/j.nuclphysb.2004.06.034}{{\em Nucl. Phys. B}
  {\bf 695} (2004) 3} [\href{http://arXiv.org/abs/hep-ex/0404008}{{\tt
  arXiv:hep-ex/0404008}}].

\bibitem{Alexa:2013xxa}
{\bf H1} collaboration, C.~Alexa {\em et.~al.}, {\it {Elastic and
  Proton-Dissociative Photoproduction of $J/\psi$ Mesons at HERA}},
  \href{http://dx.doi.org/10.1140/epjc/s10052-013-2466-y}{{\em Eur. Phys. J. C}
  {\bf 73} (2013)~no.~6 2466} [\href{http://arXiv.org/abs/1304.5162}{{\tt
  arXiv:1304.5162 [hep-ex]}}].

\bibitem{Bertulani:2005ru}
C.~A. Bertulani, S.~R. Klein and J.~Nystrand, {\it {Physics of ultra-peripheral
  nuclear collisions}},
  \href{http://dx.doi.org/10.1146/annurev.nucl.55.090704.151526}{{\em Ann. Rev.
  Nucl. Part. Sci.} {\bf 55} (2005) 271}
  [\href{http://arXiv.org/abs/nucl-ex/0502005}{{\tt arXiv:nucl-ex/0502005}}].

\bibitem{Aaij:2014iea}
{\bf LHCb} collaboration, R.~Aaij {\em et.~al.}, {\it {Updated measurements of
  exclusive $J/\psi$ and $\psi$(2S) production cross-sections in pp collisions
  at $\sqrt{s}=7$ TeV}},
  \href{http://dx.doi.org/10.1088/0954-3899/41/5/055002}{{\em J. Phys. G} {\bf
  41} (2014) 055002} [\href{http://arXiv.org/abs/1401.3288}{{\tt
  arXiv:1401.3288 [hep-ex]}}].

\bibitem{Acharya:2018jua}
{\bf ALICE} collaboration, S.~Acharya {\em et.~al.}, {\it {Energy dependence of
  exclusive $\mathrm {J}/\psi $ photoproduction off protons in ultra-peripheral
  p\textendash{}Pb collisions at $\sqrt{s_{\mathrm {\scriptscriptstyle NN}}} =
  5.02$ TeV}},  \href{http://dx.doi.org/10.1140/epjc/s10052-019-6816-2}{{\em
  Eur. Phys. J. C} {\bf 79} (2019)~no.~5 402}
  [\href{http://arXiv.org/abs/1809.03235}{{\tt arXiv:1809.03235 [nucl-ex]}}].

\bibitem{Li:2020ntd}
{\bf LHCb} collaboration, H.~Li, {\it {Z production in pPb collisions and
  charmonium production in PbPb ultra-peripheral collisions at LHCb}},
  \href{http://dx.doi.org/10.1016/j.nuclphysa.2020.121902}{{\em Nucl. Phys. A}
  {\bf 1005} (2021) 121902} [\href{http://arXiv.org/abs/2002.01863}{{\tt
  arXiv:2002.01863 [nucl-ex]}}].

\bibitem{Khachatryan:2016qhq}
{\bf CMS} collaboration, V.~Khachatryan {\em et.~al.}, {\it {Coherent $J/\psi$
  photoproduction in ultra-peripheral PbPb collisions at $\sqrt {s_{NN}} =$
  2.76 TeV with the CMS experiment}},
  \href{http://dx.doi.org/10.1016/j.physletb.2017.07.001}{{\em Phys. Lett. B}
  {\bf 772} (2017) 489} [\href{http://arXiv.org/abs/1605.06966}{{\tt
  arXiv:1605.06966 [nucl-ex]}}].

\bibitem{Acharya:2019vlb}
{\bf ALICE} collaboration, S.~Acharya {\em et.~al.}, {\it {Coherent J/$\psi$
  photoproduction at forward rapidity in ultra-peripheral Pb-Pb collisions at
  $\sqrt{s_{\rm{NN}}}=5.02$ TeV}},
  \href{http://dx.doi.org/10.1016/j.physletb.2019.134926}{{\em Phys. Lett. B}
  {\bf 798} (2019) 134926} [\href{http://arXiv.org/abs/1904.06272}{{\tt
  arXiv:1904.06272 [nucl-ex]}}].

\bibitem{Balitsky:2008zza}
I.~Balitsky and G.~A. Chirilli, {\it {Next-to-leading order evolution of color
  dipoles}},  \href{http://dx.doi.org/10.1103/PhysRevD.77.014019}{{\em Phys.
  Rev. D} {\bf 77} (2008) 014019} [\href{http://arXiv.org/abs/0710.4330}{{\tt
  arXiv:0710.4330 [hep-ph]}}].

\bibitem{Kovner:2013ona}
A.~Kovner, M.~Lublinsky and Y.~Mulian, {\it {Jalilian-Marian, Iancu, McLerran,
  Weigert, Leonidov, Kovner evolution at next to leading order}},
  \href{http://dx.doi.org/10.1103/PhysRevD.89.061704}{{\em Phys. Rev. D} {\bf
  89} (2014)~no.~6 061704} [\href{http://arXiv.org/abs/1310.0378}{{\tt
  arXiv:1310.0378 [hep-ph]}}].

\bibitem{Balitsky:2013fea}
I.~Balitsky and G.~A. Chirilli, {\it {Rapidity evolution of Wilson lines at the
  next-to-leading order}},
  \href{http://dx.doi.org/10.1103/PhysRevD.88.111501}{{\em Phys. Rev. D} {\bf
  88} (2013) 111501} [\href{http://arXiv.org/abs/1309.7644}{{\tt
  arXiv:1309.7644 [hep-ph]}}].

\bibitem{Lappi:2015fma}
T.~Lappi and H.~M\"antysaari, {\it {Direct numerical solution of the coordinate
  space Balitsky-Kovchegov equation at next to leading order}},
  \href{http://dx.doi.org/10.1103/PhysRevD.91.074016}{{\em Phys. Rev. D} {\bf
  91} (2015)~no.~7 074016} [\href{http://arXiv.org/abs/1502.02400}{{\tt
  arXiv:1502.02400 [hep-ph]}}].

\bibitem{Lappi:2016fmu}
T.~Lappi and H.~M\"antysaari, {\it {Next-to-leading order Balitsky-Kovchegov
  equation with resummation}},
  \href{http://dx.doi.org/10.1103/PhysRevD.93.094004}{{\em Phys. Rev. D} {\bf
  93} (2016)~no.~9 094004} [\href{http://arXiv.org/abs/1601.06598}{{\tt
  arXiv:1601.06598 [hep-ph]}}].

\bibitem{Dumitru:2020gla}
A.~Dumitru and R.~Paatelainen, {\it {Sub-femtometer scale color charge
  fluctuations in a proton made of three quarks and a gluon}},
  \href{http://dx.doi.org/10.1103/PhysRevD.103.034026}{{\em Phys. Rev. D} {\bf
  103} (2021)~no.~3 034026} [\href{http://arXiv.org/abs/2010.11245}{{\tt
  arXiv:2010.11245 [hep-ph]}}].

\bibitem{Dumitru:2021tvw}
A.~Dumitru, H.~M\"antysaari and R.~Paatelainen, {\it {Color charge correlations
  in the proton at NLO: Beyond geometry based intuition}},
  \href{http://dx.doi.org/10.1016/j.physletb.2021.136560}{{\em Phys. Lett. B}
  {\bf 820} (2021) 136560} [\href{http://arXiv.org/abs/2103.11682}{{\tt
  arXiv:2103.11682 [hep-ph]}}].

\bibitem{Hanninen:2017ddy}
H.~H\"anninen, T.~Lappi and R.~Paatelainen, {\it {One-loop corrections to light
  cone wave functions: the dipole picture DIS cross section}},
  \href{http://dx.doi.org/10.1016/j.aop.2018.04.015}{{\em Annals Phys.} {\bf
  393} (2018) 358} [\href{http://arXiv.org/abs/1711.08207}{{\tt
  arXiv:1711.08207 [hep-ph]}}].

\bibitem{Beuf:2016wdz}
G.~Beuf, {\it {Dipole factorization for DIS at NLO: Loop correction to the
  $\gamma^*_{T,L}\to q\overline q$ light-front wave functions}},
  \href{http://dx.doi.org/10.1103/PhysRevD.94.054016}{{\em Phys. Rev. D} {\bf
  94} (2016)~no.~5 054016} [\href{http://arXiv.org/abs/1606.00777}{{\tt
  arXiv:1606.00777 [hep-ph]}}].

\bibitem{Lappi:2016oup}
T.~Lappi and R.~Paatelainen, {\it {The one loop gluon emission light cone wave
  function}},  \href{http://dx.doi.org/10.1016/j.aop.2017.02.002}{{\em Annals
  Phys.} {\bf 379} (2017) 34} [\href{http://arXiv.org/abs/1611.00497}{{\tt
  arXiv:1611.00497 [hep-ph]}}].

\bibitem{Ducloue:2017ftk}
B.~Duclou\'e, H.~H\"anninen, T.~Lappi and Y.~Zhu, {\it {Deep inelastic
  scattering in the dipole picture at next-to-leading order}},
  \href{http://dx.doi.org/10.1103/PhysRevD.96.094017}{{\em Phys. Rev. D} {\bf
  96} (2017)~no.~9 094017} [\href{http://arXiv.org/abs/1708.07328}{{\tt
  arXiv:1708.07328 [hep-ph]}}].

\bibitem{Beuf:2021qqa}
G.~Beuf, T.~Lappi and R.~Paatelainen, {\it {Massive quarks in NLO dipole
  factorization for DIS: Longitudinal photon}},
  \href{http://dx.doi.org/10.1103/PhysRevD.104.056032}{{\em Phys. Rev. D} {\bf
  104} (2021)~no.~5 056032} [\href{http://arXiv.org/abs/2103.14549}{{\tt
  arXiv:2103.14549 [hep-ph]}}].

\bibitem{Boussarie:2016bkq}
R.~Boussarie, A.~V. Grabovsky, D.~Y. Ivanov, L.~Szymanowski and S.~Wallon, {\it
  {Next-to-Leading Order Computation of Exclusive Diffractive Light Vector
  Meson Production in a Saturation Framework}},
  \href{http://dx.doi.org/10.1103/PhysRevLett.119.072002}{{\em Phys. Rev.
  Lett.} {\bf 119} (2017)~no.~7 072002}
  [\href{http://arXiv.org/abs/1612.08026}{{\tt arXiv:1612.08026 [hep-ph]}}].

\bibitem{Escobedo:2019bxn}
M.~A. Escobedo and T.~Lappi, {\it {Dipole picture and the nonrelativistic
  expansion}},  \href{http://dx.doi.org/10.1103/PhysRevD.101.034030}{{\em Phys.
  Rev. D} {\bf 101} (2020)~no.~3 034030}
  [\href{http://arXiv.org/abs/1911.01136}{{\tt arXiv:1911.01136 [hep-ph]}}].

\bibitem{Ducloue:2017dit}
B.~Duclou\'e, E.~Iancu, T.~Lappi, A.~H. Mueller, G.~Soyez, D.~N.
  Triantafyllopoulos and Y.~Zhu, {\it {Use of a running coupling in the NLO
  calculation of forward hadron production}},
  \href{http://dx.doi.org/10.1103/PhysRevD.97.054020}{{\em Phys. Rev. D} {\bf
  97} (2018)~no.~5 054020} [\href{http://arXiv.org/abs/1712.07480}{{\tt
  arXiv:1712.07480 [hep-ph]}}].

\bibitem{Ducloue:2016shw}
B.~Duclou\'e, T.~Lappi and Y.~Zhu, {\it {Single inclusive forward hadron
  production at next-to-leading order}},
  \href{http://dx.doi.org/10.1103/PhysRevD.93.114016}{{\em Phys. Rev. D} {\bf
  93} (2016)~no.~11 114016} [\href{http://arXiv.org/abs/1604.00225}{{\tt
  arXiv:1604.00225 [hep-ph]}}].

\bibitem{Stasto:2013cha}
A.~M. Stasto, B.-W. Xiao and D.~Zaslavsky, {\it {Towards the Test of Saturation
  Physics Beyond Leading Logarithm}},
  \href{http://dx.doi.org/10.1103/PhysRevLett.112.012302}{{\em Phys. Rev.
  Lett.} {\bf 112} (2014)~no.~1 012302}
  [\href{http://arXiv.org/abs/1307.4057}{{\tt arXiv:1307.4057 [hep-ph]}}].

\bibitem{Chirilli:2011km}
G.~A. Chirilli, B.-W. Xiao and F.~Yuan, {\it {One-loop Factorization for
  Inclusive Hadron Production in $pA$ Collisions in the Saturation Formalism}},
   \href{http://dx.doi.org/10.1103/PhysRevLett.108.122301}{{\em Phys. Rev.
  Lett.} {\bf 108} (2012) 122301} [\href{http://arXiv.org/abs/1112.1061}{{\tt
  arXiv:1112.1061 [hep-ph]}}].

\bibitem{Chirilli:2012jd}
G.~A. Chirilli, B.-W. Xiao and F.~Yuan, {\it {Inclusive Hadron Productions in
  pA Collisions}},  \href{http://dx.doi.org/10.1103/PhysRevD.86.054005}{{\em
  Phys. Rev. D} {\bf 86} (2012) 054005}
  [\href{http://arXiv.org/abs/1203.6139}{{\tt arXiv:1203.6139 [hep-ph]}}].

\bibitem{Altinoluk:2014eka}
T.~Altinoluk, N.~Armesto, G.~Beuf, A.~Kovner and M.~Lublinsky, {\it
  {Single-inclusive particle production in proton-nucleus collisions at
  next-to-leading order in the hybrid formalism}},
  \href{http://dx.doi.org/10.1103/PhysRevD.91.094016}{{\em Phys. Rev. D} {\bf
  91} (2015)~no.~9 094016} [\href{http://arXiv.org/abs/1411.2869}{{\tt
  arXiv:1411.2869 [hep-ph]}}].

\bibitem{Watanabe:2015tja}
K.~Watanabe, B.-W. Xiao, F.~Yuan and D.~Zaslavsky, {\it {Implementing the exact
  kinematical constraint in the saturation formalism}},
  \href{http://dx.doi.org/10.1103/PhysRevD.92.034026}{{\em Phys. Rev. D} {\bf
  92} (2015)~no.~3 034026} [\href{http://arXiv.org/abs/1505.05183}{{\tt
  arXiv:1505.05183 [hep-ph]}}].

\bibitem{Iancu:2016vyg}
E.~Iancu, A.~H. Mueller and D.~N. Triantafyllopoulos, {\it {CGC factorization
  for forward particle production in proton-nucleus collisions at
  next-to-leading order}},
  \href{http://dx.doi.org/10.1007/JHEP12(2016)041}{{\em JHEP} {\bf 12} (2016)
  041} [\href{http://arXiv.org/abs/1608.05293}{{\tt arXiv:1608.05293
  [hep-ph]}}].

\bibitem{Bodwin:2007fz}
G.~T. Bodwin, H.~S. Chung, D.~Kang, J.~Lee and C.~Yu, {\it {Improved
  determination of color-singlet nonrelativistic QCD matrix elements for S-wave
  charmonium}},  \href{http://dx.doi.org/10.1103/PhysRevD.77.094017}{{\em Phys.
  Rev. D} {\bf 77} (2008) 094017} [\href{http://arXiv.org/abs/0710.0994}{{\tt
  arXiv:0710.0994 [hep-ph]}}].

\bibitem{Berger:2012wx}
J.~Berger and A.~M. Stasto, {\it {Exclusive vector meson production and small-x
  evolution}},  \href{http://dx.doi.org/10.1007/JHEP01(2013)001}{{\em JHEP}
  {\bf 01} (2013) 001} [\href{http://arXiv.org/abs/1205.2037}{{\tt
  arXiv:1205.2037 [hep-ph]}}].

\bibitem{Berger:2010sh}
J.~Berger and A.~Stasto, {\it {Numerical solution of the nonlinear evolution
  equation at small x with impact parameter and beyond the LL approximation}},
  \href{http://dx.doi.org/10.1103/PhysRevD.83.034015}{{\em Phys. Rev. D} {\bf
  83} (2011) 034015} [\href{http://arXiv.org/abs/1010.0671}{{\tt
  arXiv:1010.0671 [hep-ph]}}].

\bibitem{Bendova:2019psy}
D.~Bendova, J.~Cepila, J.~G. Contreras and M.~Matas, {\it {Solution to the
  Balitsky-Kovchegov equation with the collinearly improved kernel including
  impact-parameter dependence}},
  \href{http://dx.doi.org/10.1103/PhysRevD.100.054015}{{\em Phys. Rev. D} {\bf
  100} (2019)~no.~5 054015} [\href{http://arXiv.org/abs/1907.12123}{{\tt
  arXiv:1907.12123 [hep-ph]}}].

\bibitem{Beuf:2020dxl}
G.~Beuf, H.~H\"anninen, T.~Lappi and H.~M\"antysaari, {\it {Color Glass
  Condensate at next-to-leading order meets HERA data}},
  \href{http://dx.doi.org/10.1103/PhysRevD.102.074028}{{\em Phys. Rev. D} {\bf
  102} (2020) 074028} [\href{http://arXiv.org/abs/2007.01645}{{\tt
  arXiv:2007.01645 [hep-ph]}}].

\bibitem{heikki_mantysaari_2020_4229269}
G.~Beuf, H.~Hänninen, T.~Lappi and H.~Mäntysaari, {\it Color glass condensate
  at next-to-leading order meets hera data (software)},  2020.
\newblock \url{https://doi.org/10.5281/zenodo.4229269}.

\bibitem{Mantysaari:2020lhf}
H.~M\"antysaari, K.~Roy, F.~Salazar and B.~Schenke, {\it {Gluon imaging using
  azimuthal correlations in diffractive scattering at the Electron-Ion
  Collider}},  \href{http://dx.doi.org/10.1103/PhysRevD.103.094026}{{\em Phys.
  Rev. D} {\bf 103} (2021)~no.~9 094026}
  [\href{http://arXiv.org/abs/2011.02464}{{\tt arXiv:2011.02464 [hep-ph]}}].

\bibitem{Lappi:2020ufv}
T.~Lappi, H.~M\"antysaari and J.~Penttala, {\it {Relativistic corrections to
  the vector meson light front wave function}},
  \href{http://dx.doi.org/10.1103/PhysRevD.102.054020}{{\em Phys. Rev. D} {\bf
  102} (2020)~no.~5 054020} [\href{http://arXiv.org/abs/2006.02830}{{\tt
  arXiv:2006.02830 [hep-ph]}}].

\bibitem{Iancu:2015vea}
E.~Iancu, J.~D. Madrigal, A.~H. Mueller, G.~Soyez and D.~N. Triantafyllopoulos,
  {\it {Resumming double logarithms in the QCD evolution of color dipoles}},
  \href{http://dx.doi.org/10.1016/j.physletb.2015.03.068}{{\em Phys. Lett. B}
  {\bf 744} (2015) 293} [\href{http://arXiv.org/abs/1502.05642}{{\tt
  arXiv:1502.05642 [hep-ph]}}].

\bibitem{Iancu:2015joa}
E.~Iancu, J.~D. Madrigal, A.~H. Mueller, G.~Soyez and D.~N. Triantafyllopoulos,
  {\it {Collinearly-improved BK evolution meets the HERA data}},
  \href{http://dx.doi.org/10.1016/j.physletb.2015.09.071}{{\em Phys. Lett. B}
  {\bf 750} (2015) 643} [\href{http://arXiv.org/abs/1507.03651}{{\tt
  arXiv:1507.03651 [hep-ph]}}].

\bibitem{Beuf:2014uia}
G.~Beuf, {\it {Improving the kinematics for low-$x$ QCD evolution equations in
  coordinate space}},  \href{http://dx.doi.org/10.1103/PhysRevD.89.074039}{{\em
  Phys. Rev. D} {\bf 89} (2014)~no.~7 074039}
  [\href{http://arXiv.org/abs/1401.0313}{{\tt arXiv:1401.0313 [hep-ph]}}].

\bibitem{Ducloue:2019ezk}
B.~Duclou\'e, E.~Iancu, A.~H. Mueller, G.~Soyez and D.~N. Triantafyllopoulos,
  {\it {Non-linear evolution in QCD at high-energy beyond leading order}},
  \href{http://dx.doi.org/10.1007/JHEP04(2019)081}{{\em JHEP} {\bf 04} (2019)
  081} [\href{http://arXiv.org/abs/1902.06637}{{\tt arXiv:1902.06637
  [hep-ph]}}].

\bibitem{Hoodbhoy:1996zg}
P.~Hoodbhoy, {\it {Wave function corrections and off forward gluon
  distributions in diffractive $J/\mathrm{\psi}$ electroproduction}},
  \href{http://dx.doi.org/10.1103/PhysRevD.56.388}{{\em Phys. Rev. D} {\bf 56}
  (1997) 388} [\href{http://arXiv.org/abs/hep-ph/9611207}{{\tt
  arXiv:hep-ph/9611207}}].

\end{thebibliography}\endgroup

\clearpage

\appendix

\onecolumngrid
\section{Full expressions}
In this Appendix we show the full expressions for the vector meson production cross section at next-to-leading order that are too cumbersome to be included in the Letter. For completeness, we also include here explicitly the virtual photon and vector meson wave functions at NLO that are necessary inputs for our calculation and available in the literature.

Throughout this Appendix, we use the notation where $\xt_{ij}=\xt_i-\xt_j$ and $\overline Q^2 = z_0(1-z_0)Q^2$. The photon or quarkonium plus momentum fractions carried by the quark, antiquark and the gluon are $z_0,z_1$ and $z_2$, respectively. We work in the frame where the photon and quarkonium plus momentum $q^+$ is large.

\subsection{Longitudinal photon wave function}
\label{appendix:photon}
The virtual photon wave function with massive quarks is derived in Ref.~\cite{Beuf:2021qqa} and included here for reference in the case where the photon has no transverse momentum. All expressions are shown in the conventional dimensional regularization (CDR) scheme. For the $\gamma \to q\bar q$ splitting, the leading order wave function in $D$ dimensions reads
\begin{equation}
	\label{eq:photon_qq_LO}
	\Psi_{\gamma^*\,\mathrm{LO}}^{q\bar q} = \delta_{\alpha_0 \alpha_1}  \cdot \frac{-e e_f Q}{2 \pi q^+} \sqrt{z_0(1-z_0)} \bar u(0) \gamma^+ v(1) \left(\frac{\sqrt{\overline Q^2+m_q^2}}{2\pi|\xt_{01}|} \right)^{D/2-2} K_{D/2-2}\left( |\xt_{01}| \sqrt{\overline Q^2+m_q^2} \right).
\end{equation}
Here $u(0) = u(k_0, h_0)$ is the spinor describing a quark with momentum $k_0$, color $\alpha_0$ and helicity $h_0$, and similarly $v(1)$ is the antiquark spinor, $h_1$ the antiquark helicity and $\alpha_1$ the antiquark color. 
Note that the wave function normalization here differs from that of Ref.~\cite{Beuf:2021qqa} by a factor $(2 q^+ \sqrt{z_0 z_1})^{-1}$
as the phase space measure $\der^2 \xt \frac{\der z}{4\pi}$  used in this work differs from that of Ref.~\cite{Beuf:2021qqa}.

At next-to-leading order one adds a loop correction, and finds 
\begin{equation}
	\label{eq:photon_qq_NLO}
	\Psi_{\gamma^*\,\mathrm{NLO}}^{q \bar q} = \delta_{\alpha_0 \alpha_1}  \cdot \frac{-e e_f Q}{2 \pi q^+} z_0(1-z_0) \bar u(0) \gamma^+ v(1) \left(\frac{\alpha_s C_F}{2\pi} \right) \tilde \vcal^L + \Psi_{h.f.}.
\end{equation}
Here $\Psi_{h.f}.$ refers to the helicity flip component of the longitudinal photon wave function.  As discussed in the Letter, it does not contribute to the cross section at this order in $\as$ and $v^2$ and can be dropped. The explicit expression for $\tilde \vcal^L$ is shown below.

The wave function describing $\gamma^* \to q\bar q g$ splitting is obtained by calculating a tree level gluon emission contribution. The resulting wave function reads
\begin{multline}
	\label{eq:photon_qqg_NLO}
		\Psi_{\gamma^*}^{ q \bar q g} = t^a_{\alpha_0 \alpha_1} \frac{4e e_f Q g }{2q^+} \epsilon_\sigma^{j*}\sqrt{\frac{z_0z_1}{z_2}}
		\Bigg\{ \bar u(0) \gamma^+\left[ \left( 1+ \frac{z_2}{2z_0}\right) \delta^{ij}- \frac{z_2}{4z_0} [\gamma^i, \gamma^j] \right]v(1) \mathcal{I}_{(f)}^i \\
		- \bar u(0) \gamma^+\left[ \left( 1+ \frac{z_2}{2z_1}\right) \delta^{ij}+ \frac{z_2}{4z_1} [\gamma^i, \gamma^j] \right]v(1) \mathcal{I}_{(g)}^i 
		-\frac{m_q}{2} \left[ \frac{z_0}{z_0+z_2} \left(\frac{z_2}{z_0}\right)^2\mathcal{I}_{(f)}-\frac{z_1}{z_1+z_2} \left(\frac{z_2}{z_1}\right)^2\mathcal{I}_{(g)} \right] \bar u(0)\gamma^+ \gamma^j v(1).
		\Bigg\}.
\end{multline}
Here $a$ is the gluon color and $\sigma$ refers to the gluon polarization.
This again differs from the result presented in Ref.~\cite{Beuf:2021qqa} by a factor $(2q^+ \sqrt{z_0z_1z_2})^{-1}$ due to the different phase space measure as discussed above.

Let us next present detailed expressions for the special functions that are used in the NLO wave functions above and used when we present  the final scattering amplitude for the exclusive vector meson production in Appendix~\ref{appendix:nlo_xs}.
First, in the $\Psi_{\gamma^*\,\mathrm{NLO}}^{q \bar q}$, Eq.~\eqref{eq:photon_qq_NLO}, one has
\begin{multline}
	\label{eq:V_L_NLO}
	\tilde \vcal^L =  \left\{  \left[\frac{3}{2}+\log(\frac{\alpha}{z_0})+\log(\frac{\alpha}{1-z_0}) \right] \left[ \frac{(4\pi)^{2-D/2}}{2-D/2}\Gamma\left(3-\frac{D}{2}\right)+\log(\frac{\xt_{01}^2\mu^2}{4})+2\gamma_E\right]+\frac{1}{2}\right\} \\
	\cdot \left( \frac{\sqrt{\overline Q^2+m_q^2}}{2\pi|\xt_{01}|} \right)^{D/2-2} K_{D/2-2} \left(|\xt_{01}| \sqrt{\overline Q^2+m_q^2} \right) \\
	 + \left[  \log^2 \left(\frac{z_0}{1-z_0} \right)-\frac{\pi^2}{3}+\frac{5}{2}+\Omega_\vcal(\gamma;z_0) + L(\gamma;z_0) \right] K_0 \left(|\xt_{01}| \sqrt{\overline Q^2+m_q^2}\right) + \mathcal{\tilde I}_\vcal (z_0,\xt_{01}).
\end{multline}
Here $\gamma_E$ is the Euler's constant and $\mu^2$ the  scale in dimensional regularization which eventually cancels,
and the infrared regulator $\alpha$ discussed also in Sec.~\ref{sec:nlo_vm} regulates the soft infrared divergence in loop integrals.
Furthermore, the following definitions have been used
\begin{multline}
	\label{eq:Omega}
	\Omega_\vcal(\gamma;z) = \frac{1}{2z} \left[ \log(1-z) + \gamma \log(\frac{1+\gamma}{1+\gamma-2z}) \right]+\frac{1}{2(1-z)} \left[ \log(z) + \gamma \log(\frac{1+\gamma}{1+\gamma-2(1-z)}) \right] \\
	+\frac{1}{4z(1-z)} (\gamma-1) \log(\frac{\overline Q^2+m_q^2}{m_q^2}) +\frac{m_q^2}{2\overline Q^2} \log(\frac{\overline Q^2+m_q^2}{m_q^2})
\end{multline}
with $\gamma = \sqrt{1+\frac{4m_q^2}{Q^2}}$.
Additionally 
\begin{equation}
	\label{eq:Lgammaz}
	L(\gamma;z) = \li \left(\frac{1}{1-\frac{1}{2z}(1-\gamma)}\right)+\li \left(\frac{1}{1-\frac{1}{2(1-z)}(1-\gamma)}\right)+\li \left(\frac{1}{1-\frac{1}{2z}(1+\gamma)}\right)+\li \left(\frac{1}{1-\frac{1}{2(1-z)}(1+\gamma)}\right),
\end{equation}
where $\li$ is the dilogarithm function.
The function $\mathcal{\tilde I}_\vcal(z,\xt_{01})$ is defined as 
\begin{equation}
\label{eq:Iv}
    \mathcal{\tilde I}_\vcal(z,\xt_{01}) =\mathcal{\tilde I}_{\vcal_{(a)+(b)}} (z,\xt_{01}) + \mathcal{\tilde I}_{\vcal_{(c)+(d)}}(z,\xt_{01}),
    \end{equation} 
    with 
\begin{multline}
	\label{eq:J1}
	\mathcal{\tilde I}_{\vcal_{(a)+(b)}}(z,\xt_{01})  = \int_0^1 \frac{\dd[]{\xi}}{\xi} \left(- \frac{2\log \xi}{1-\xi}+\frac{1+\xi}{2} \right) \left[ 2K_0 \left(|\xt_{01}| \sqrt{\overline Q^2+m_q^2}\right) 
	 -K_0 \left(|\xt_{01}| \sqrt{\overline Q^2+m_q^2+\frac{(1-z)\xi}{1-\xi}m_q^2}\right)
	  \right. \\
	  - \left. K_0 \left(|\xt_{01}| \sqrt{\overline Q^2+m_q^2+\frac{z\xi}{1-\xi}m_q^2}\right)
		\right]
\end{multline}
and
\begin{multline}
	\label{eq:J2}
	\mathcal{\tilde I}_{\vcal_{(c)+(d)}}(z,\xt_{01}) = m_q^2 \int_0^1 \dd{\xi} \int_0^1 \dd{x} \\
	 \Bigg\{ \left[ K_0 \left(|\xt_{01}| \sqrt{\overline Q^2+m_q^2}\right)-K_0 \left(|\xt_{01}| \sqrt{\frac{\overline{Q}^2+m_q^2}{1-x}+\kappa}\right)\right]
	\frac{C^L_m}{(1-\xi)(1-x)\left[x(1-\xi)+\frac{\xi}{1-z}\right]\left[\frac{x(\overline Q^2+m_q^2)}{1-x}+\kappa\right]}   \\
	+ (z \to 1-z) \Bigg\}.
\end{multline}
The coefficients above read
\begin{align}
	\label{eq:kappa} 
	\kappa &= \frac{\xi m_q^2}{(1-\xi)(1-x)\left[x(1-\xi)+\frac{\xi}{1-z}\right]} \left[\xi(1-x)+x \left(1-\frac{z(1-\xi)}{1-z}\right) \right] \quad\text{and} \\
	\label{eq:Clm}
	C^L_m &= \frac{z^2 (1-\xi)}{1-z} \left[ -\xi^2 +x(1-\xi) \frac{1+(1-\xi) \left(1+\frac{z \xi}{1-z}\right)}{x(1-\xi) + \frac{\xi}{1-z}} \right]. 
\end{align}

For the $\Psi_{\gamma^*}^{q \bar q g}$ wave function \eqref{eq:photon_qqg_NLO} we also need the following special functions:
\begin{equation}
\label{eq:Ifg}
\begin{split}
\mathcal{I}^{i}_{(f)} &= \mathcal{I}^{i}(\xt_{0+2;1},\xt_{20},\overline{Q}^2_{(f)},\omega_{(f)},\lambda_{(f)}), \quad\quad \mathcal{I}_{(f)} = \mathcal{I}(\xt_{0+2;1},\xt_{20},\overline{Q}^2_{(f)},\omega_{(f)},\lambda_{(f)})\\
\mathcal{I}^{i}_{(g)} &= \mathcal{I}^{i}(\xt_{0;1+2},\xt_{21},\overline{Q}^2_{(g)},\omega_{(g)},\lambda_{(g)}),\quad\quad \mathcal{I}_{(g)} = \mathcal{I}(\xt_{0;1+2},\xt_{21},\overline{Q}^2_{(g)},\omega_{(g)},\lambda_{(g)}).
\end{split}
\end{equation}
Here, the integrals are defined as
\begin{align} 
	\label{eq:Icomp}
	\mathcal{I}^i(\bt,\rt,\overline Q^2, \omega, \lambda) &=\frac{ i \mu^{2-D/2}}{ 2(4\pi)^{D-2}}  \rt^i \int_0^\infty \dd[]{u} u^{1-D/2} e^{-u(\overline Q^2+m_q^2)} e^{-\bt^2/(4u)} \int_0^{u/\omega} \dd[]{t} t^{-D/2} e^{-t \omega \lambda m_q^2} e^{-\rt^2/(4t)} \\
	\label{eq:I}
	\mathcal{I}(\bt,\rt,\overline Q^2, \omega, \lambda) &= \frac{ \mu^{2-D/2}}{ (4\pi)^{D-2}} \int_0^\infty \dd[]{u} u^{1-D/2} e^{-u(\overline Q^2+m_q^2)} e^{-\bt^2/(4u)} \int_0^{u/\omega} \dd[]{t} t^{1-D/2} e^{-t \omega \lambda m_q^2} e^{-\rt^2/(4t)}
\end{align}
with the definitions
\begin{align}
	\label{eq:omega_f}
	\omega_{(f)} &= \frac{z_0 z_2}{z_1(z_0+z_2)^2} \\ %\quad &\omega_g &= \frac{z_1 z_2}{z_0(z_1+z_2)^2} \\
	\overline Q^2_{(f)} &= z_1 (1-z_1) Q^2 \\ %\quad &\overline Q^2_{(g)} &= z_0 (1-z_0) Q^2 \\
	\lambda_{(f)} &= \frac{z_1 z_2}{z_0} \\ %\quad &\lambda_{(g)} &= \frac{z_0 z_2}{z_1} \\
	\xt_{n+m;p} &= - \xt_{p;n+m} = \frac{z_n \xt_n+z_m \xt_m}{z_n+z_m} - \xt_p.
\end{align}
Additionally, $\omega_{(g)}$, $\overline{Q}_{(g)}^2$ and $\lambda_{(g)}$ are obtained from $\omega_{(f)}$, $\overline{Q}_{(f)}^2$ and $\lambda_{(f)}$ by exchanging $z_0 \leftrightarrow z_1$.

\subsection{Vector meson wave function}
\label{appendix:vm}

The vector meson wave function is written in Eq.~\eqref{eq:NR_expansion} in terms of the coefficients $C^k_{n \leftarrow m}$. These coefficients have been calculated in Ref.~\cite{Escobedo:2019bxn}, and are shown in this Appendix in the case of the longitudinal polarization. The quark and antiquark helicities at the level of the leading-order wave function $\phi^{q\bar q}$ are $h_0'$ and $h_1'$. In the full quarkonium wave function $\Psi_V$ they are denoted by $h_0$ and $h_1$. In the nonrelativistic limit ($k=0$) we have $C^0_{n\leftarrow m}=0$ for all $n\neq m$ at leading order, and the only non-zero coefficient is 
\begin{equation}
    C^0_{q\bar q \leftarrow q\bar q} = \frac{4\pi}{\sqrt{\nc}} \delta_{\alpha_0 \alpha_1} \delta_{h_0 h_0'} \delta_{h_1 h_1'} \delta\left(z_0-\frac{1}{2}\right).
\end{equation}
Here $\alpha_0$ and $\alpha_1$ are the quark and antiquark colors, respectively. At next-to-leading order in $\as$, the non-zero coefficients are
\begin{multline}
	\label{eq:C0_qq}
		C^{0}_{q \bar q \leftarrow q \bar q} = \frac{1}{\sqrt{N_c}} \delta_{\alpha_0 \alpha_1} \delta_{h_0 h_0'}\delta_{h_1 h_1'} \Bigg\{ 4\pi \delta\left(z_0-\frac{1}{2}\right)(1+\delta Z) 
		+\frac{\alpha_s C_F}{2\pi}\frac{16\pi z_0(1-z_0)}{(z_0-1/2)^2} \left[\theta\left(z_0-\frac{1}{2}-\alpha\right)+ \theta\left(-z_0+\frac{1}{2}-\alpha\right) \right] \\
		 \times \Bigg\{K_0(\tau) +\left[ \theta \left(z_0-\frac{1}{2}\right)(1-z_0)+\theta\left(\frac{1}{2}-z_0\right)z_0\right]
		\left[ \frac{(z_0-1/2)^2-1/2}{z_0(1-z_0)}\left( K_0(\tau)-\frac{\tau}{2}K_1(\tau)\right)-\frac{(z_0-1/2)^2}{2z_0(1-z_0)}\tau K_1(\tau) \right] \Bigg\} \Bigg\}
\end{multline}
with $\tau = 2m_q |\xt_{01}| |z_0-1/2|$, and 
\begin{multline}
	\label{eq:C0_qqg}
		C_{q \bar q g \leftarrow q \bar q}^{0} = \frac{1}{\sqrt{N_c}}\frac{2m_q}{\sqrt{q^+ k_{0'}^+}} \frac{g}{2\pi} \delta_{h_1 h_1'}t^a_{\alpha_0 \alpha_1} \epsilon_\sigma^{j*} \left( \frac{2m_qz_2}{2\pi |\xt_{20}| \mu}\right)^{D/2-2}\cdot 4\pi \delta\left(z_1-\frac{1}{2}\right)\\
		\times 	\Bigg[ -i(1-z_2)\sqrt{\frac{2z_2}{1-2z_2}}\frac{\xt_{20}^i}{ |\xt_{20}|} K_{D/2-1}\big(2m_qz_2 |\xt_{20}|\big) \bar u(0) \left[\delta^{ij}-\frac{z_2}{2(1-z_2)} [\gamma^i,\gamma^j]\right]\gamma^+ u(0')  \\
		+z_2\sqrt{\frac{2z_2}{1-2z_2}} K_{D/2-2}\big(2m_qz_2 |\xt_{20}|\big) \bar u(0) \gamma^j \gamma^+ u(0')\Bigg] + \text{antiquark contribution} 
\end{multline}
Here $\alpha$ is the infrared cutoff discussed in Sec.~\ref{sec:nlo_vm} which forces $z_2>\alpha$, and $a$ is the color of the gluon with polarization $\sigma$. 
This coefficient describes the process where a nonrelativistic quark with spinor $u(0')$ (the plus momentum of the nonrelativistic quark is $k_{0'}^{+} = q^+/2$) emits a gluon and becomes relativistic, described by the spinor $\bar u(0)$.
Compared to Ref.~\cite{Escobedo:2019bxn}, the factor $m_q/\sqrt{q^+ k_{0'}^+}$ comes from the fact that in Ref.~\cite{Escobedo:2019bxn} the rest frame spinors were used for the quarks.  Also, the integration variable in~\cite{Escobedo:2019bxn} is $z_2/2$ instead of $z_2$, which explains the factor $2$ in front.
The antiquark contribution gives eventually exactly the same contribution as the gluon emission from the quark shown above.

In the loop correction the wave function renormalization coefficient is needed:
\begin{multline}
	\label{eq:delta_Z}
	\delta Z = -\frac{\alpha_s C_F}{2\pi} \left[ \frac{1}{D-4}(4\log(2\alpha)+3)+2\log(2\alpha) \left(\log(\frac{m_q^2}{4\pi\mu^2})+1+\log(2\alpha)\right) \right. \\
	\left. +(4\log(2\alpha)+3)\frac{\gamma_E}{2}+\frac{3}{2}\log(\frac{m_q^2}{4\pi\mu^2})-2\right].
\end{multline}
Here $\mu^2$ is the mass scale used in dimensional regularization, and it cancels in the final result. It should be noted that the above expression for the wave function renormalization depends on the scheme for the quark mass renormalization. Although the procedure for the mass renormalization differs in Refs.~\cite{Beuf:2021qqa} and~\cite{Escobedo:2019bxn}, ultimately the mass scheme is identical for both the photon and meson wave functions and therefore they can be combined consistently.

In addition to the NLO corrections we consider the relativistic corrections. A systematic way to include relativistic corrections at leading order in $\alpha_s$ and order $v^k$ in velocity is given by Eq.~\eqref{eq:NR_expansion_alphas0} where only terms with $k_1+k_2+k_3 \leq k$ are kept. In general, this involves nonperturbative constants $\phi^{q \bar q}_{h_0 h_1}(k_1,k_2,k_3)$ defined in Eq.~\eqref{eq:phi-constants} that can be calculated using an ansatz for the leading order wave function $\phi^{q \bar q}$. In Ref.~\cite{Lappi:2020ufv} relativistic expansion for   $\phi^{q \bar q}$ around the origin in terms of the derivatives of the rest frame wave function was derived. The wave function $\phi^{q \bar q}$ was calculated at order $v^2$, and for the longitudinal polarization it reads:
\begin{equation}
    \label{eq:NRQCD_expansion}
    \phi^{q \bar q}_{h_0 h_1} = \delta_{h_0, -h_1}\pi \sqrt{\frac{2}{m_q}} \phi_\text{RF}(0) \left[\delta\left(z_0-\frac{1}{2}\right)+\frac{1}{6 m_q^2} \frac{\nabla^2 \phi_\text{RF}(0)}{\phi_\text{RF}(0)} \Bigg( \left(\frac{5}{2}+\rt^2 m_q^2  \right)\delta\left(z_0-\frac{1}{2}\right) -\frac{1}{4} \partial_{z_0}^2\delta\left(z_0-\frac{1}{2}\right) \Bigg)  \right] ,
\end{equation}
where $\phi_\text{RF}(\vec r)$ is the rest frame wave function in coordinate space. Using this wave function one gets Eq.~\eqref{eq:v^2_LFWF}.

\subsection{Exclusive vector meson production cross section at next-to-leading order}\label{appendix:nlo_xs}
The final result for the exclusive vector meson production amplitude at next-to-leading order is shown in Eq.~\eqref{eq:total_NLO}. It can be written in terms of the special functions and definitions appearing in the NLO  virtual photon and vector meson wave functions shown explicitly in Appendices~\ref{appendix:photon} and~\ref{appendix:vm}. The kernel corresponding to the virtual correction in Eq.~\eqref{eq:total_NLO} reads
	\begin{multline}
    \label{eq:K_qq_NLO}
        \kcal_{q \bar q}^\nlo(\Ydip) = 
         \left[ \kcal+
        \mathcal{\tilde I}_\nu\left(z=\frac{1}{2},\xt_{01}\right)
        +K_0(\zeta) \left(6-\frac{\pi^2}{3}+\Omega_\vcal\left(\gamma;z=\frac{1}{2}\right) + L\left(\gamma;z=\frac{1}{2}\right)-3\log(\frac{|\xt_{01}|m}{2})-3\gamma_E \right) \right]\\
        \times N_{01}(\Ydip).
\end{multline} 
Here $\mathcal{\tilde I}_\vcal$ is the integral shown in Eq.~\eqref{eq:Iv}, $\Omega_\vcal$ is given in Eq.~\eqref{eq:Omega} and $L$ in Eq.~\eqref{eq:Lgammaz}.
We have denoted $\zeta =|\xt_{01}|\sqrt{\frac{1}{4}Q^2+m_q^2}$, and
\begin{multline}
	\label{eq:K0}
		\kcal = \int_0^{1/2} \dd[]{z_0} \Bigg\{ 16z_0(1-z_0)K_0\left(|\xt_{01}|\sqrt{\overline Q^2+m_q^2}\right)2z_0 \left[K_0(\tau)-\tau K_1(\tau) \right] \\
		+ \frac{1}{(z_0-1/2)^2}\Bigg\{ 
			16z_0(1-z_0)K_0\left(|\xt_{01}|\sqrt{ \overline Q^2+m_q^2}\right) \left[
				2z_0(1-z_0)K_0(\tau)-z_0\left(
					K_0(\tau)-\frac{\tau}{2}K_1(\tau)
				\right)
			\right] \\
			- K_0(\zeta) \left[
				1+2\left(z_0-\frac{1}{2}\right) \left[
					1+2 \gamma_E + 2\log(m_q|\xt_{01}|)+2\log\left(\frac{1}{2}-z_0\right)
				 \right]
			\right]
		\Bigg\}\Bigg\}.
\end{multline}
The real gluon emission contribution  reads
\begin{multline}
    \label{eq:K_qqg}
        \kcal_{q \bar q g}(\Yqqg) = -32 \pi m_q  \Bigg\{ \frac{i \xt_{20}^i}{|\xt_{20}|} K_1(2m_q z_2 |\xt_{20}|) \left[\left( (1-z_2)^2 + z_2^2 \right) \ical_{(f)}^i + (2z_2^2-1)(1-2z_2) \ical_{(g)}^i \right] N_{012}(\Yqqg) \\
		+4m_q z_2^3 K_0(2m_q z_2|\xt_{20}|) \left[ \ical_{(f)} - \frac{1-2z_2}{1+2z_2} \ical_{(g)} \right] N_{012}(\Yqqg) \\
		+\frac{1}{8\pi^2} \left((1-z_2)^2+z_2^2\right) \frac{1}{m_q z_2 |\xt_{20}|^2} K_0(\zeta) e^{-\xt_{20}^2\B} N_{01}(\Yqqg)
		\Bigg\}.
\end{multline}
The integrals $\ical_{(f)}$ and $\ical_{(g)}$ together with $\ical_{(f)}^i$ and $\ical_{(g)}^i$ are given in Eqs.~\eqref{eq:Ifg}, \eqref{eq:Icomp} and \eqref{eq:I}.
We emphasize that all expressions here are finite given that the lower limit for the $z_2$ integral is non-zero, as required in physically meaningful kinematics, see Eq.~\eqref{eq:z2min}.

\end{document}